\documentclass[11pt]{article}

\usepackage[margin=0.9in]{geometry}

\usepackage{comment}
\usepackage{graphicx, color}
\usepackage{mathtools}          %.. extends amsmath
\usepackage{amssymb,amsmath}
\usepackage{bm,bbm}
\usepackage{float}
\usepackage[usenames,dvipsnames]{xcolor}
\usepackage{appendix}
\usepackage{url}
\usepackage{marginnote}
\usepackage{natbib}
\usepackage[normalem]{ulem}  % in preamble
\usepackage{authblk}

   \newcommand{\vct}[1]  {\ensuremath{\boldsymbol{#1}}}    %.. bold italic
 \newcommand{\vu} {\vct{u}}

 \newcommand{\vF} {\vct{F}}

 \newcommand{\vJ} {\vct{J}}

 \newcommand{\vS} {\vct{S}}

 \renewcommand{\d} {\mathrm{d}}
\newcommand{\squishlist}{
   \begin{list}{$\bullet$}
    { \setlength{\itemsep}{0pt}      \setlength{\parsep}{3pt}
      \setlength{\topsep}{3pt}       \setlength{\partopsep}{0pt}
      \setlength{\leftmargin}{1.5em} \setlength{\labelwidth}{1em}
      \setlength{\labelsep}{0.5em} } }

\newcommand{\squishlisttwo}{
   \begin{list}{$\bullet$}
    { \setlength{\itemsep}{0pt}    \setlength{\parsep}{0pt}
      \setlength{\topsep}{0pt}     \setlength{\partopsep}{0pt}
      \setlength{\leftmargin}{2em} \setlength{\labelwidth}{1.5em}
      \setlength{\labelsep}{0.5em} } }

\newcommand{\squishend}{
    \end{list}  }

\newcommand{\beq}{\begin{equation}}
\newcommand{\eeq}{\end{equation}}

\newcommand{\lan} {\langle}
\newcommand{\ran} {\rangle}

\newcommand{\eps}{\varepsilon}

\newcommand{\ol}  [1] {\overline{#1}}     %.. or use \bar

\newcommand{\ou}{\overline{\vu}^\ell}

\newcommand{\vSigma}{\bm{\Sigma}}
\renewcommand{\vJ}{\bm{J}}

\newcommand{\ovA}{\overline{\bm A}}
\newcommand{\ovB}{\overline{\bm B}}

\newcommand{\xx}{\bm{x}}

\newcommand{\Sbarij}{ \overline{S}^\ell_{ij}}

\newcommand{\tr}[1] 
  {\mathrm{Tr} \left\{ {#1} \right\}}
\newcommand{\purple}[1]{\textcolor{black}{#1}}

\title{Why MHD turbulence forms sheets: energy conversion and cascade geometry}

\author[1]{Damiano Capocci\thanks{Corresponding author: dcapocci@ed.ac.uk}}
\author[2]{Moritz Linkmann\thanks{Corresponding author: moritz.linkmann@ed.ac.uk}}

\affil[1]{School of Physics and Astronomy and Higgs Centre,
The University of Edinburgh,
Edinburgh EH9 3FD, United Kingdom}

\affil[2]{School of Mathematics and Maxwell Institute for Mathematical Sciences,
University of Edinburgh,
Edinburgh EH9 3FD, United Kingdom}

\date{\today}

\begin{document}

\maketitle

\begin{abstract}
\noindent Sheet-like structures are ubiquitous in magnetohydrodynamic systems and are known to play a crucial role in astrophysical plasma transport. Here, by applying a theoretical framework to data from numerical simulations, we demonstrate that current-sheet formation is a direct consequence of energy conversion between kinetic and magnetic fields. The strain--magnetic-field organisation associated with this conversion favours the emergence of sheet-like structures, and analytical calculations show that sheet-like geometry constitutes the dominant flow configuration. We further show that the same conversion-driven reorganisation of the strain motion is responsible for the depletion of the \purple{mean} Inertial energy transfer, a previously observed phenomenon whose physical origin has remained unclear. Finally, we identify experimentally accessible diagnostics through which these predictions can be tested in multi-spacecraft observations of turbulent plasmas.
\end{abstract}

% \keywords{Suggested keywords}
%.. Use showkeys class option if keyword display desired

  \maketitle

%==========================================================
%% \section{\label{sec:level1} Section title}

%\marginnote{\textbf{general motivational intro}}
Magnetohydrodynamics (MHD) is a fluid model that describes the evolution of electrically conducting fluids, such as plasmas, through the coupled dynamics of the velocity field and the magnetic field. A central feature of plasma systems is the formation of coherent structures where these fields develop strong gradients. Two important examples are velocity sheets, where the flow changes sharply across a narrow layer, and current sheets, where strong magnetic-field gradients produce foliated electric currents. Both these two structures are relevant in turbulent plasmas where they control energy dissipation and may trigger processes such as magnetic reconnection \citep{sweet1958,parker1957} or shear-flow instabilities \citep{fujimoto2017}. For this reason, understanding how velocity sheets and current sheets form is essential for explaining energy transfer and release in astrophysical, space, and laboratory plasmas.

%\marginnote{\textbf{sheet creation vs production}}
It is useful to distinguish between the \emph{production} and the \emph{creation} of sheet-like structures. The term production emphasises the processes by which sheets develop, sharpen, and are sustained by the spatiotemporal and interscale behaviour of the system, rather than merely the instant at which they first appear. By contrast, creation is a more ambiguous notion, since it may refer to many different procedures or choices of initial condition, such as introducing a magnetic field into an initially hydrodynamic flow or modifying an already magnetised MHD state. In this sense, "production" is the more precise concept when studying the physical mechanisms responsible for the persistence and evolution of velocity and current sheets.

%\marginnote{\textbf{sheet thinning}}
Recently it was shown that current-sheet thinning is not merely a geometrical feature of MHD turbulence, but appears to provide the main physical mechanism that governs the energy transfer from large to small scale \cite{capocci2025}. In this picture, the cascade is controlled primarily by the stretching and compression of current layers in regions of magnetic strain. Conversely, the energy transfer relative to pure Inertial contribution, which would dominate in the pure hydrodynamic case, is strongly depleted on average.

%\marginnote{\textbf{sheets and energy balance}}
In the MHD energy equations, one can isolate a term that describes the conversion between kinetic and magnetic energy \cite{offermans2018, rincon2019, schekochihin2004}. Such a term is dynamically constrained: its sign and mean value can be determined by the global energy balance. At the same time, it admits a direct geometrical interpretation, since it depends on the relative orientation between the magnetic field and the strain-rate tensor. This makes it possible to identify the contribution of sheet-like configurations to the energy conversion process.

%\marginnote{\textbf{Depletion of the Inertial transfer}}
Moreover, the strain--magnetic-field alignment associated with energy conversion does not only favour sheet-like structures, but also modifies the geometry of the velocity strain itself. This has direct consequences for the purely Inertial channel of the cascade, namely the transfer associated only with velocity gradient deformations. In hydrodynamic (HD) turbulence this constitutes the sole interscale energy-transfer channel and is responsible for the forward energy cascade, namely the transfer of energy from large to small scale. In MHD, instead, its mean contribution is strongly depleted as previously observed in \cite{yousef2007,servidio2008, alexakis2013,yang2021}. The geometrical mechanism responsible for this depletion, however, remains unclear.
%\footnote{\red{there is also a paper with Sean (Li, Matthaeus, Sean, Chen, PRL  (2026))}}.

%\marginnote{\textbf{synthesis/conciliation}}
The aim of this work is to connect this energy-conversion mechanism with the formation of sheet-like structures and the relative current-sheet thinning described above. We show that the formation of sheet-like structures is favoured by the conversion between kinetic and magnetic energy, while current-sheet thinning provides the mechanism through which this process is sustained and saturated in a statistically stationary state. In this way, sheet formation and current-sheet thinning are not separate phenomena, but two competing aspects of the same energy-transfer and energy balance in MHD turbulence. We also provide a geometrical explanation for the depletion of the Inertial transfer. The depletion can be interpreted as a consequence of the magnetic back-reaction: the same strain--magnetic-field organisation associated with energy conversion and sheet formation also reorganises the strain field, weakening the average purely kinetic energy transfer. Consequently, mechanisms such as vortex stretching and strain self-amplification, which sustain the hydrodynamic energy cascade, are suppressed on average in MHD.

\begin{comment}
    Equations to show:

\begin{itemize}
    \item Standard viscosity EOM.
    \item Unfiltered energy balance (potentially), people would think that the energy conversion is an a consequence of the filtering method.
    \item filtered energy balance and SGS stress tensors listed a la $B_0 \neq 0 $ paper. $\mathcal{W}$ should be used as a symbol and defined as an equation to emphasise. I don't find a reason to emphasise the spatial transport terms.
    \item list all the subfluxes.
    \item Write the RSC term via the energy-eigenvalues decomposition.
    \item Write the new identity and connect it with the sheet saturation.
    \item Depletion of the Inertial transfer: decompose the fluxes.
    
\end{itemize}
\end{comment}

\section{Theory}

In the next two sections we describe the theoretical methodology and the analytical results. The theoretical predictions are eventually tested against data from numerical simulations respectively in secs.~\ref{sec:results_sheets}--\ref{sec:results_depletion}.

\subsection{Energy balance and sheet sustainment}
\label{sec:sheet_intro}

The governing equations of three dimensional (3D) incompressible homogeneous MHD turbulence are:
\begin{align}
    &\frac{\partial\bm{u}}{\partial t} + \bm{u} \cdot \nabla \bm{u} = -\nabla \left(p + \frac{b^2}{2} \right) + \bm{b} \cdot \nabla \bm{b} + \nu \Delta \bm{u} + \bm{F} \label{eq:mom_eq} \\
    &\frac{\partial \bm{b}}{\partial t} + \bm{u} \cdot \nabla \bm{b} =  \bm{b} \cdot \nabla \bm{u} + \mu \Delta \bm{b} \label{eq:ind_eq} \\
    &\nabla \cdot \bm{u} =   0 \label{eq:divzeros_vel} \\
    &\nabla \cdot \bm{b} =   0  \label{eq:divzeros}
\end{align}
where $\bm{u}$ are the velocity fluctuations, $\bm{b}$ is the magnetic field in Alfv\'en speed units: $\bm{b}/\sqrt{4\pi\rho} \to \bm{b}$, $p$ is the pressure divided by the constant fluid density $\rho$, $\bm{F}$ is a mechanical (large scale) forcing term, $\nu$ and $\mu$ the kinematic viscosity and magnetic diffusivity respectively. Both $\bm{u}$ and $\bm{b}$ have zero mean.  the separation of scales is characterised by the forcing scale $\ell_f$ and the Kolmogorov dissipation scale $\eta=(\nu^3/\varepsilon)^{1/4}$, with $\ell_f/\eta\gg1$. 

From the above equations, we can derive the evolution equation of the kinetic and magnetic energy. 
\begin{align}
  \partial_t {E}_u
        + \nabla \cdot \vct{J}_u
      & =
        - {\cal W}
        - {\cal D}_{u} 
        + \bm{u} \cdot \bm{F},
  \label{eq:Eu-ls}
 \\
  \partial_t {E}_b
        + \nabla \cdot \vct{J}_b
      & =
         \hspace{0.27cm} {\cal W}
        - {\cal D}_{b} ,
  \label{eq:Eb-ls}
\end{align}
where we isolate the terms that can be expressed as the divergence of vectors via the definition of the energy transport currents $\bm{J}_u$ and $\bm{J}_b$. It follows that the $\mathcal{D}$ terms represent the dissipative effects. The multiscale nature of MHD can be quantified by the separation between the forcing (large) scale $L_f$ and the Kolmogorov dissipation scale $\eta=(\nu^3/\varepsilon)^{1/4}$ where $\varepsilon= \lan  \mathcal{D}_u\ran + \lan  \mathcal{D}_b\ran$ is the energy dissipation rate and the angle brackets denote the volume average. For fully developed turbulence we obtain $L_f/\eta\gg1$. Furthermore, we can observe that both the equations present a term appearing with opposite sign, that can be interpreted as the energy conversion term. Specifically, its expression reads \citep{offermans2018, rincon2019, schekochihin2004}
\begin{equation}
    \mathcal{W} = S_{ij} \, b_i b_j
    \label{eq:rsc_unf}
\end{equation}
where $S_{ij} = \left( \partial_i u_j + \partial_j u_i  \right)/2$ represents the strain rate tensor namely the symmetric part of the velocity gradient tensor. Throughout this work, the Einstein convention for repeated indices is adopted. We can easily show that, at the stationary state and by virtue of periodic boundaries or infinite domain, the energy conversion term equals, on average, the magnetic dissipation rate
\begin{equation}
    \lan \mathcal{W} \ran  = \lan \mathcal{D}_b \ran >0
    \label{eq:constr_dyn}
\end{equation}
From a geometrical point of view, this term describes the alignment between the Maxwell tensor and the strain-rate tensor eigenvectors. In this regard, the corresponding strain-rate tensor eigenvalues determine the type of deformation that the element of fluid undergoes due to strain motion. Since the tensor $S_{ij}$ is symmetric, it admits three real eigenvalues $\{\lambda_1,\lambda_2,\lambda_3\}$ where  the incompressibility of the velocity field provides the pointwise constraint $\lambda_1 + \lambda_2 + \lambda_3\ = 0$. As a consequence, the largest eigenvalue $\lambda_3$ is always non-negative and the smallest $\lambda_1$ is non-positive while $\lambda_2$ can be either positive or negative. Hence, $\lambda_i>0$ implies that along the direction of the corresponding eigenvector the element of fluid is extended, conversely $\lambda_i <0$ describes a contraction along the eigendirection. It follows that the classifier of the strain-deformation type is given by the product between the strain-rate tensor eigenvalues i.e. $ \lambda_1 \lambda_2  \lambda_3$. In presence of sheet-like structures this product is negative as we have two extensional directions and a single contractile one; conversely tubular structures present a \purple{positive} product since we have one single extensional direction and two contractile directions. Before proceeding, we underline that the sign and magnitude of the mean energy conversion term imposes an alignment between the magnetic field and the strain-motion. It is clear that this local alignment contributes to the loss of instantaneous isotropy of MHD system. In previous works, instantaneous loss of isotropy was documented, see e.g.~\cite{muller2003,matthaeus2012}, in MHD with no mean magnetic field.
%which is attributed essentially to the formation of a local non-zero mean magnetic field
To the Authors' knowledge and capability, the strain-magnetic field alignment due to energy conservation is the most immediate way of identifying this property. We extend the present discussion around fig.~\ref{fig:depletion_scheme} (b).

To identify the role of sheet-structures in the energy conversion term, we express the above energy conversion term in eq.~\eqref{eq:rsc_unf} via the strain/magnetic field alignment:
\begin{equation}
    \mathcal{W} =  |\bm{b}|^2   \lambda_i \cos^2{\theta_{\bm{b},i}}
    \label{eq:conv_spec}
\end{equation}
where $\theta_{{\bm{b}},i}$ is the angle between the magnetic field and the strain-rate eigenvector associated with the eigenvalue $\lambda_i$. The energy conversion is therefore governed by the magnetic-field strength, the individual strain eigenvalues and alignment angles, as well as by their collective alignment-weighted contribution, through which the different eigendirections can reinforce or cancel one another. Notably, a sheet-like velocity structure, given by $\lambda_2 >0$, can provide a positive energy conversion meaning that the energy is transferred from the kinetic to the magnetic energy. This naturally raises the question of whether such configurations are statistically favoured in MHD. To address this, we consider the mean strain-eigenvalue product for which it is possible to derive the following identity
\begin{comment}
    \begin{equation}
    \lan \lambda_1 \lambda_2  \lambda_3 \ran = - 4 \lan \bm{\nabla u}^T \,   \bm{J}\, \bm{\Sigma}^T \ran  - \dfrac{\nu}{4} \lan \bm{\nabla\nabla u} \bm{\nabla\nabla u}^T  \ran  -  \dfrac{\eta}{4} \lan \bm{\nabla} \bm{\nabla b} \bm{\nabla\nabla b}^T  \ran 
\end{equation}
%the following formula, in matrix notation is horrible
\end{comment}
\begin{equation}
    \lan \lambda_1 \lambda_2  \lambda_3 \ran =   \lan \partial_j u_i J_{jk} \Sigma_{ki} \ran  - \dfrac{\nu}{4} \lan \partial_k \partial_j u_i \, \partial_k \partial_j u_i \ran  -  \dfrac{\mu}{4} \lan \partial_k \partial_j b_i \, \partial_k \partial_j b_i \ran 
    \label{eq:eig_prod_eq}
\end{equation}
where $\Sigma_{ij} = \left( \partial_i b_j + \partial_j b_i  \right)/2$ and $J_{ij} = \left( \partial_i b_j - \partial_j b_i  \right)/2$ are respectively the symmetric and antisymmetric parts of the magnetic field gradient tensor which represents the current-free and current carrying spatial variations of the same magnetic field. The former is the magnetic analogue of the strain-rate tensor, while the latter is the magnetic analogue of the antisymmetric, vorticity-carrying part of the velocity gradient tensor namely $\Omega_{ij} = \left( \partial_i u_j - \partial_j u_i  \right)/2$. To derive eq.~\eqref{eq:eig_prod_eq}, we made use of the solenoidality of both magnetic and velocity field alongside the lack of solid boundaries. Details of the derivation can be found in \cite{thesis}.

In eq.~\eqref{eq:eig_prod_eq}, it is evident that the two last terms are  negative by construction. They can be interpreted as gradient-dissipation terms as they are proportional to the diffusivity and magnetic resistivity. This indicates that the turbulent dissipative effects promote the formation of sheets as they give a negative contribution to the RHS. By contrast, the sign of the first term on the RHS cannot be inferred a priori. Remarkably, this gradients-contraction term was previously interpreted as a \emph{current-sheet thinning} mechanism and shown to provide the dominant contribution to the interscale transfer in MHD turbulence \cite{capocci2025,capocci2026}. Moreover, the previous investigation showed that this term is positive and therefore counteracts the sheet formation promoted by the other two terms in eq.~\eqref{eq:eig_prod_eq}. However, it can be shown that in fully developed turbulence, the first term is smaller in magnitude than the sign-definite gradient-dissipation terms \cite{thesis}, leading to an overall $ \lan \lambda_1 \lambda_2  \lambda_3 \ran <0$ which indicate that sheet-like structure are dominant. The echo of the turbulent cascade does not end with this term; indeed, even the product of the strain-rate tensor eigenvalues plays a role in the interscale transfer, a connection that forms the central focus of sec.~\ref{sec:results_depletion}.

From a physical point of view, the role of current-sheet thinning in the formation of velocity sheets  acts as a saturation mechanism. If current-sheet thinning promotes the formation of such structures, the resulting increase in strain would further enhance the thinning of the current-sheet layer, leading to a potentially unbounded self-reinforcing mechanism. 
In this sense, the mechanism is reminiscent of Lenz’s law, to the extent that the response generated by the system opposes the process driving its further evolution.

%This is especially relevant in the presence of quantities that are bounded from above, such as gradient dissipation and current-sheet thinning. 

%It is important to note a previous study \citep{seta2020} already analysed the magnetic-energy conversion in eq.~\eqref{eq:rsc_unf} through its decomposition in the strain eigenframe and its modification by Lorentz-force back-reaction. Their analysis, however, did not exploit this decomposition to identify which strain geometries preferentially realise positive conversion, as done here through eq.~\eqref{eq:conv_spec}. Building on this, we connect the geometry selected by positive conversion, through the magnetic-gradient dynamics, to current-sheet thinning and interscale energy transfer.

\purple{A previous study has already analysed the magnetic-energy conversion in eq.~\eqref{eq:rsc_unf} through its decomposition in the strain eigenframe and has also characterised the morphology of the resulting magnetic structures \citep{seta2020}. Here, we use an equivalent eigenframe representation, making explicit both the strain eigenvalues and the strain--magnetic-field alignment angles, to address a different question: which local velocity geometries preferentially realise positive kinetic-to-magnetic conversion?} Building on this, we connect the geometry selected by positive conversion, through the magnetic-gradient dynamics, to current-sheet thinning and interscale energy transfer.
\begin{figure}
	\begin{center}
         %\noindent\makebox[\textwidth]{
         \includegraphics[width=0.8\columnwidth]{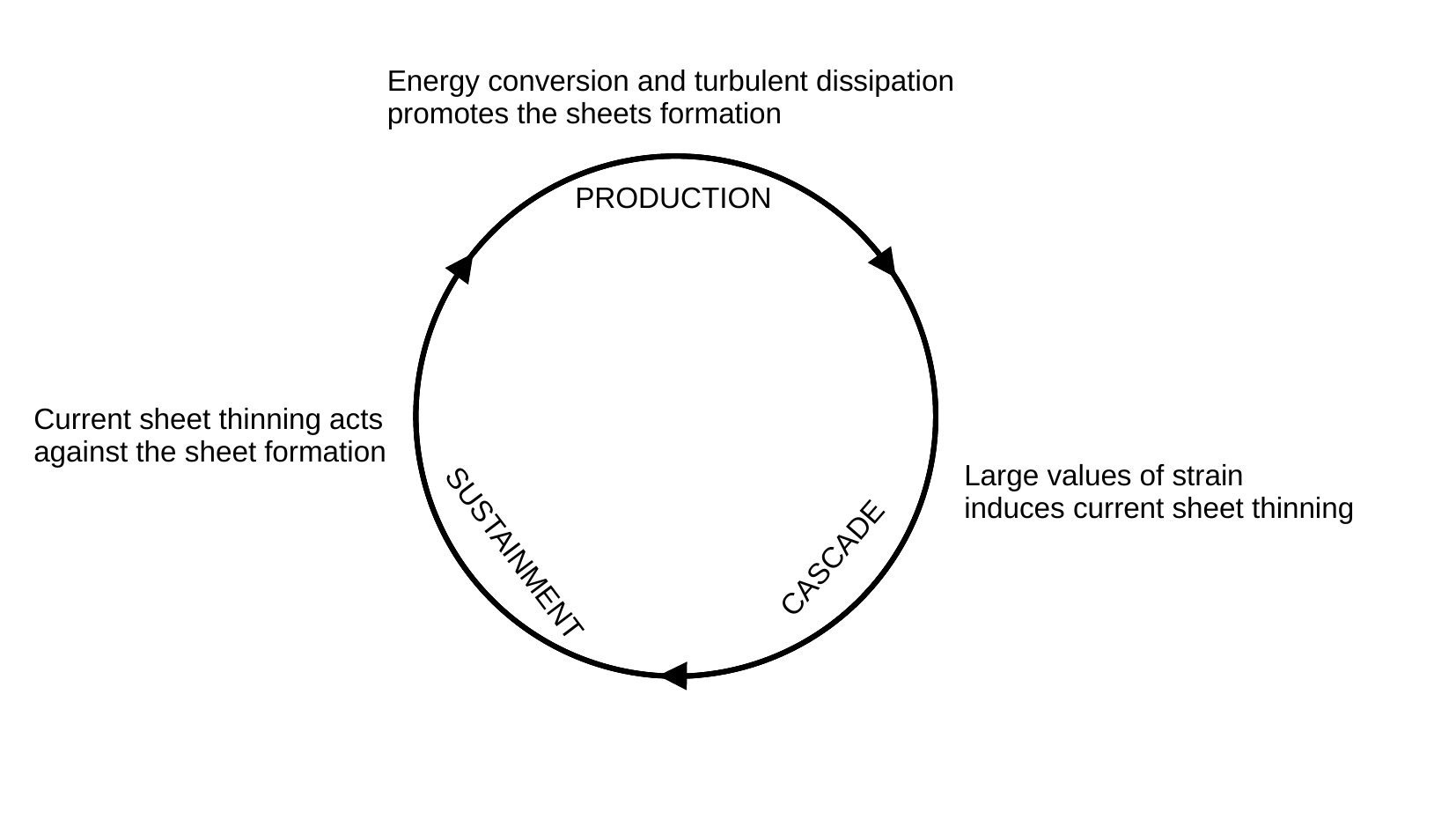} 
         %}
    \end{center}
	 \caption{
Scheme of the velocity and current structures mutual sustainment. 
}
\label{fig:sketch_scheme}
\end{figure}

\subsection{Energy transfer and sheet sustainment}

%\marginnote{\textbf{Why filtering?}}
%The previous discussion shows that the strain geometry is statistically constrained by its alignment with the magnetic field through the energy conversion. At the same time, the product of the strain-rate eigenvalues is related to a balance of gradient contractions.
The appearance of the current-sheet-thinning contraction in eq.~\eqref{eq:eig_prod_eq}, dominating the interscale energy transfer in MHD turbulence, therefore suggests a connection between the scale-by-scale organisation of the strain field and the transfer of energy across scales. To expose this scale-by-scale connection, we introduce a coarse-grained formulation that disentangles the contributions arising at different scales. The corresponding coarse-grained energy balances then allow us to distinguish between the conversion of kinetic and magnetic energy at the resolved scale and the transfer of energy across that scale.

For an arbitrary scalar field $f$ over a domain $V$,  we define the filtered field at scale $\ell$ as
\begin{equation}
\ol{f}^{\ell}(\xx)
=
\int_V G^\ell(\bm{r}) \, f(\xx+\bm{r}) \, \mathrm{d}^3 r ,
\label{eq:coarse_gr}
\end{equation}
where $G^\ell$ is the filter kernel \citep{germano1992}, which must satisfy a number of minor properties like evenness, \emph{fast} decay and continuity. The resolved kinetic and magnetic energies are then defined as $E^\ell_u = \overline{u}^\ell_i \overline{u}_i^\ell/2$ and $E^\ell_b = \overline{b}^\ell_i \overline{b}_i^\ell/2$. More generally and independently of the employed filter type, the application of the filtering procedure to the MHD equations gives the coarse-grained energy balances \citep{offermans2018}
\begin{align}
  \partial_t {E}^\ell_u
        + \nabla \cdot \vct{{\cal J}}^\ell_u
      & =
        - \Pi^{I,\ell} - \Pi^{M,\ell}
        - {\cal W}^\ell
        - {\cal D}^\ell_{u} 
        + \ou \cdot \ol{\vF}^\ell,
  \label{eq:Eu-ls_filt}
 \\
  \partial_t {E}^\ell_b
        + \nabla \cdot \vct{{\cal J}}^\ell_b
      & =
        - \Pi^{A,\ell} - \Pi^{D,\ell}
        + {\cal W}^\ell
        - {\cal D}^\ell_{b} ,
  \label{eq:Eb-ls_filt}
\end{align}
which retain the same structure as their unfiltered counterparts viz eqs.~\eqref{eq:Eu-ls}--\eqref{eq:Eb-ls},  now restricted to quantities resolved at scale $\ell$, with the addition of the energy flux terms $\Pi^\ell$s. The flux superscripts are labelled according to the nonlinear terms of the MHD equations from which they originate: Inertial ($I$), Maxwell ($M$), Dynamo ($D$), and Advection ($A$); see \cite{capocci2025} for further details.
%I understand that this is not great, we might put the non linear expression in brackets
Regarding the flux sign, we adopt the notation for which a $\Pi^\ell>0$ corresponds to an energy transfer from the scale $\ell$ to the subfilter, hence smaller, scales. In turbulence theory, the range of scales over which the total energy flux is approximately constant and equal to the mean dissipation rate is referred to as the \emph{inertial range}. 

Besides the interscale fluxes, the coarse-grained balances contain the resolved-scale kinetic-to-magnetic energy conversion
%the prev sentence is brute in the connection
\begin{equation}
{\cal W}^\ell = \Sbarij \,\overline{b}^\ell_i \, \overline{b}^\ell_j 
\label{eq:RSC}
\end{equation}
which describes the energy conversion at scale $\ell$, as opposed to the energy fluxes, describing transfer across $\ell$. Following \cite{aluie2010}, the stationary-state constraint in eq.~\eqref{eq:constr_dyn} extends to the resolved conversion, yielding $\langle\mathcal{W}^\ell\rangle>0$.

%\marginnote{\textbf{Why single/multiscale subfluxes splitting - IMPORTANT}} 
We now return to the central question of this section: how the organisation of the resolved strain at scale $\ell$ is related to the turbulent cascade across that same scale. To make this comparison explicit, we separate the fluxes into single-scale and multiscale contributions. This distinction is essential because the strain geometry is defined at the filtering scale $\ell$, whereas the interscale transfer also contains interactions involving smaller scales. This separation can be, remarkably, carried out exactly for a Gaussian filter, $G^\ell(r)\sim\exp(-\bm r^2/2\ell^2)/\ell^3$, following the methodology of \cite{johnson2020,johnson2021}. The same formalism also expresses the fluxes in terms of gradients of the filtered fields. Defining the velocity- and magnetic-gradient tensors as $\overline{A}^\ell_{ij}=\partial_j\overline{u}^\ell_i$ and $\overline{B}^\ell_{ij}=\partial_j\overline{b}^\ell_i$, respectively, the four energy fluxes take the form \citep{capocci2025}
\begin{align}
    \Pi^{I,\ell}
    &= -\ell^2 \tr{(\ovA^\ell)^t \ovA^\ell (\ovA^\ell)^t}
    -\int_0^{\ell^2}\d\theta \,
    \tr{(\ovA^\ell)^t
    \left[\overline{\ovA^{\sqrt{\theta}}(\ovA^{\sqrt{\theta}})^t}^{\phi}
    -\ovA^\ell(\ovA^\ell)^t \right]}
    \label{eq:Pi-I-exact} \\
    \Pi^{M,\ell} &= \ell^2 \tr{(\ovA^\ell)^t \ovB^\ell (\ovB^\ell)^t} +\int_0^{\ell^2}\d\theta\, \tr{(\ovA^\ell)^t \left[ \overline{\ovB^{\sqrt{\theta}}(\ovB^{\sqrt{\theta}})^t}^{\,\phi} -\ovB^\ell(\ovB^\ell)^t \right]} \label{eq:Pi-M-exact} \\ 
    \Pi^{A,\ell} &= -\ell^2 \tr{(\ovB^\ell)^t \ovB^\ell (\ovA^\ell)^t} -\int_0^{\ell^2}\d\theta\, \tr{(\ovB^\ell)^t \left[ \overline{\ovB^{\sqrt{\theta}}(\ovA^{\sqrt{\theta}})^t}^{\,\phi} -\ovB^\ell(\ovA^\ell)^t \right]} \label{eq:Pi-A-exact} \\ 
    \Pi^{D,\ell} &= \ell^2 \tr{(\ovB^\ell)^t \ovA^\ell (\ovB^\ell)^t} +\int_0^{\ell^2}\d\theta\, \tr{(\ovB^\ell)^t \left[ \overline{\ovA^{\sqrt{\theta}}(\ovB^{\sqrt{\theta}})^t}^{\,\phi} -\ovA^\ell(\ovB^\ell)^t \right]}. \label{eq:Pi-D-exact} 
\end{align}
%\marginnote{\textbf{Why single/multiscale subfluxes splitting}}
\purple{where $\phi = \sqrt{\ell^2 - \theta}$}. In each expression, the first term is the single-scale contribution since it is expressed entirely by gradients evaluated at scale $\ell$. The integral term is instead the multiscale contribution as it accounts for the coupling between the $\ell$ and smaller scales. As intended, the single-scale contribution isolates the part of the energy transfer generated by the same resolved gradients that encode the local flow geometry and mutual fields alignments at scale $\ell$. The multiscale contribution, while retaining its own geometrical content, quantifies the coupling between gradients at scale $\ell$ with the hierarchy of subfilter scales. %This geometrical tendency is meaningful at that scale without invoking the full hierarchy of smaller scales. 
%The single-scale part of the flux then identifies the contribution of the cascade generated by that same resolved geometry, while the multiscale part measures how the balance at $\ell$ is modified by coupling to subfilter scales.
%Sheet-like structure formation should be regarded as a scale-by-scale geometrical process, whereas the cascade describes how these scale-dependent geometries are dynamically coupled. 

%\marginnote{\textbf{Advantages of this method}}
To expose the physical processes contained in these expressions, we decompose the velocity and magnetic-field gradients into their symmetric and antisymmetric parts. The resulting exact expansion expresses each energy flux in terms of contractions among the velocity strain, rotation rate, magnetic strain and electric current density, that were introduced above, in sec.~\ref{sec:sheet_intro}. Each subflux can therefore be associated with a physically interpretable process whose strength is determined by the local alignment of the gradients at each scale. %From this perspective, sheet formation can be regarded as a scale-by-scale geometrical process, while the cascade describes how these scale-dependent geometries are dynamically coupled. The single-scale contribution isolates the energy transfer generated by the geometry resolved at $\ell$, whereas the multiscale contribution measures how this transfer is modified by interactions with smaller scales. 
To identify each subflux arising from this decomposition of eqs.~\eqref{eq:Pi-I-exact}--\eqref{eq:Pi-D-exact}, we make use of the notation
\begin{equation}
    \Pi^{X,\ell}_{ABC} = \Pi^{X,\ell}_{s, ABC} + \Pi^{X,\ell}_{m, ABC}
    \label{eq:notation}
\end{equation}
where $X$ is the flux identifier, $A,B,$ and $C$ are the tensors arising from the gradient tensors decomposition while $s$ and $m$ denote the single and multiscale parts of the RHS of eqs.~\eqref{eq:Pi-I-exact}--\eqref{eq:Pi-D-exact}. See Appendix A in \cite{capocci2025} for a more general discussion.

%\marginnote{\textbf{Connection with sheets formation - ideas}}
In this regard, the relevance of the single-scale contribution becomes apparent after decomposing the velocity and magnetic gradient tensors into their symmetric and antisymmetric parts, namely specialising the tensors $A$, $B$ and $C$ entering eq.~\eqref{eq:notation} . In particular, the single-scale Maxwell flux, i.e. the first term in eq.~\eqref{eq:Pi-M-exact}, contains the contribution
\begin{equation}
   \Pi^{M,\ell}_{s,S J \Sigma}    
     = \;
      2 \ell^2 \, \tr{\big( \overline{\vS}^\ell \big)^t \,
       \overline{\vJ}^\ell \big(\overline{\vSigma}^\ell\big)^t } 
          \label{eq:Pi-M-current-sheet}  
\end{equation}
which, apart from the scale-dependent prefactor, can be identified through the cyclic property of the trace with the same gradient contraction appearing in the fluxes of eqs.~\eqref{eq:Pi-A-exact}--\eqref{eq:Pi-D-exact}. As mentioned above, previous works \cite{capocci2025,capocci2026}, associated this term with the physical mechanism of the current-sheet thinning and its back-reaction on the velocity field. \purple{Furthermore, its quantification in homogeneous MHD turbulence, both without and with a strong mean magnetic field, showed that $\langle \Pi^{M,\ell}_{s,S J\Sigma} \rangle >0$ at all scales.}
%Furthermore, its quantification in both isotropic and strongly anisotropic homogeneous MHD turbulence showed that $\langle \Pi^{M,\ell}_{s,S J\Sigma} \rangle >0$ at every scale. 
Together with its multiscale counterpart, this mechanism provides the dominant contribution to the mean energy transfer, accounting for approximately $80\%$ of the total.

%\marginnote{\textbf{Connection with sheets formation - formulae}}
Here the central result of this section emerges: the geometry of sheet formation and the dominant mechanism of the MHD cascade are not merely related, but are encoded in the same gradient contraction appearing in eq.~\eqref{eq:eig_prod_eq}. Its positive mean competes with the negative-definite gradient-dissipation terms, which favour a negative eigenvalue product and hence sheet-like deformation. The forward cascade therefore does not merely accompany sheet formation: through the back-reaction associated with current-sheet thinning, it provides the saturation mechanism that prevents \emph{unbounded} sheet amplification. Sheet formation and current-sheet thinning thus emerge as two sides of the same dynamical balance, allowing sheet-like structures to be sustained in a statistically stationary turbulent state, as described by the scheme in fig.~\ref{fig:sketch_scheme}. In conclusion, the left-hand side of eq.~\eqref{eq:eig_prod_eq} is itself closely connected to an energy-flux contribution, a point we return to in the next section.

\begin{comment}
\begin{equation}
    \lan b_i b_j S_{ij} \ran = \int_0^\infty \Pi^{M,\ell}_{s,SJ\Sigma} \d \ell^2  + \int_0^\infty  \lan   \overline{b}^{\ell}_i \overline{b}^{\ell}_j \overline{S}^{\ell}_{ij,kk} \ran    \d \ell^2
\end{equation}
\end{comment}

\subsection{Depletion of the Inertial transfer}
\label{sec:depl_inertial_theory}

%Earlier in sec.~\ref{sec:sheet_intro}, we showed that energy conversion imposes a preferred strain--magnetic-field organisation and favours sheet-like geometry. The previous section then connected this geometry to the dominant mechanism of the MHD cascade, current-sheet thinning, which provides the saturation of sheets formation and originate both from the Lorentz force and the induction equation non-linearity. 

\begin{figure}
	\begin{center}
         %\noindent\makebox[\textwidth]{
         \includegraphics[width=0.49\columnwidth]{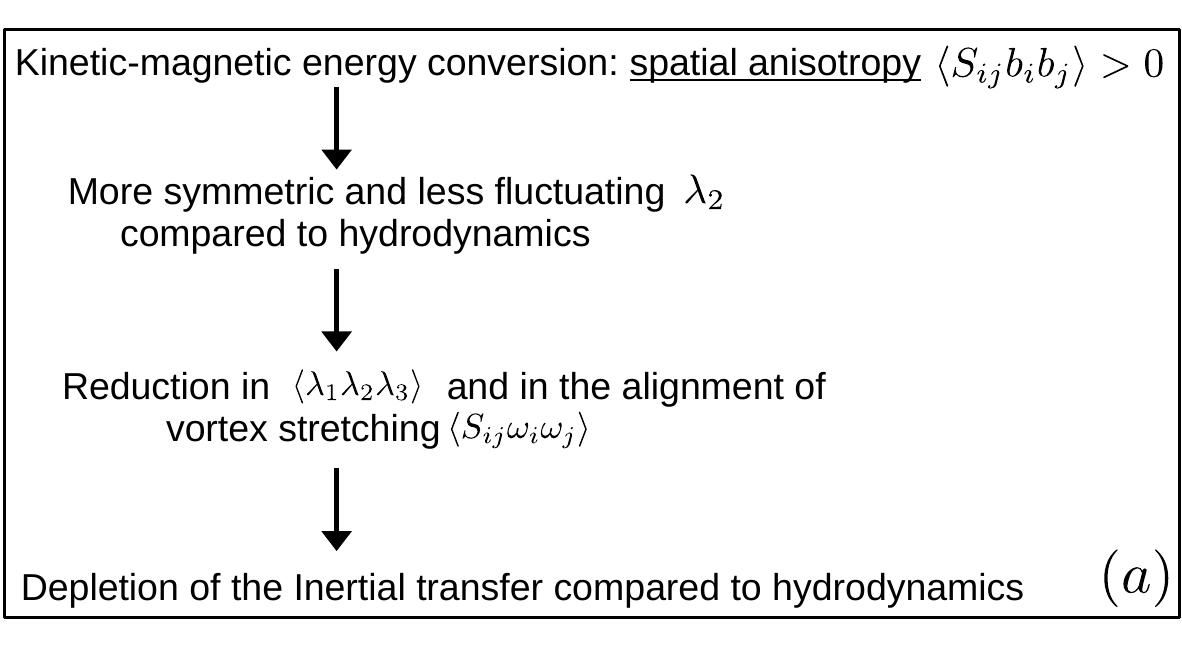} 
         %\includegraphics[width=0.65\columnwidth]{pictures/scheme_cascade.png} 
         %}
        \includegraphics[width=0.49\columnwidth]{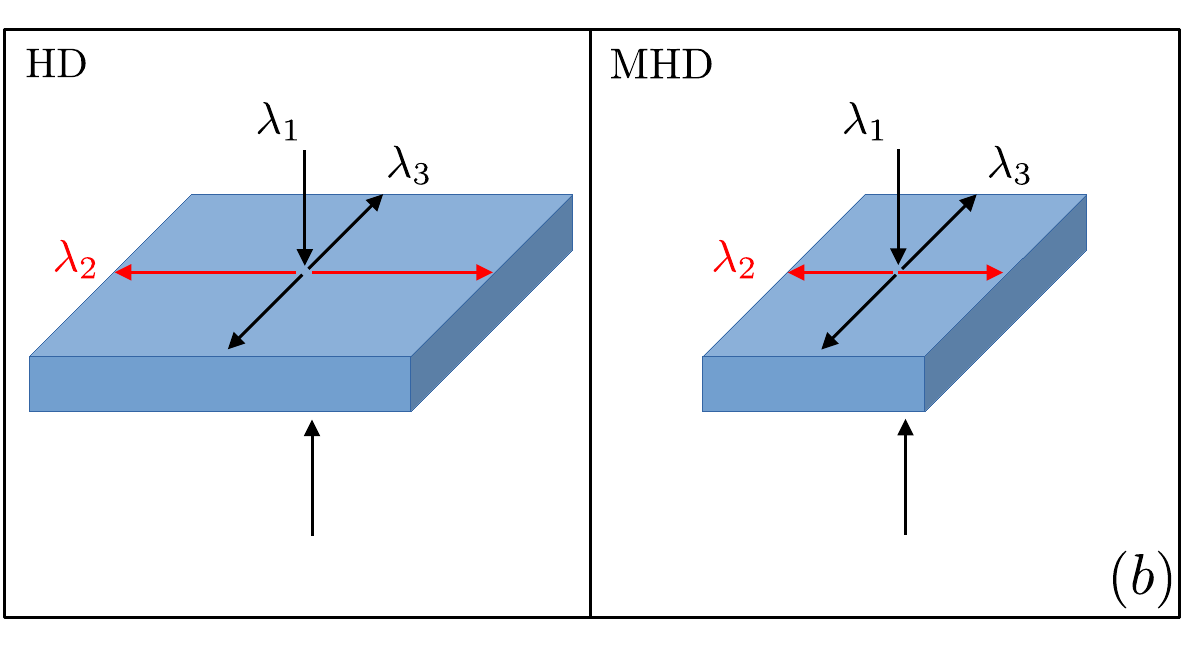}
    \end{center}
	 \caption{
Left panel, scheme of the Inertial transfer depletion which is relative to the \emph{cascade} branch of the scheme in fig.~\ref{fig:sketch_scheme}. Right panel, local element of fluid deformation behind sheet-like structures. In MHD we see that the anisotropy is indeed the anisotropy of the extensional direction strength making MHD sheet-like structures more similar to lasagne/ribbons rather than sheets. }
%the intensity of the contraction is a bit reduced as well, the lasagna-like structure is truly a good representation, this comparison is not original though, see Perry's seminar. }
\label{fig:depletion_scheme}
\end{figure}

We now turn to the opposite side of the same cascade organisation: the Inertial transfer. In this regard, previous studies \cite{yousef2007,servidio2008, alexakis2013,yang2021} showed that the purely kinetic energy transfer is strongly depleted in MHD compared with HD. More recently, \cite{capocci2025,capocci2026} traced this depletion to a suppression of the main mechanisms driving the forward Inertial transfer in HD turbulence, namely strain self-amplification and vortex stretching.  In light of the previous sections, the question that arises spontaneously is whether the strain--magnetic-field anisotropy imposed by the energy conservation can explain this depletion. More specifically, can the constraints in eqs.~\eqref{eq:rsc_unf}-\eqref{eq:constr_dyn} affect the mutual alignment of the gradients entering $\Pi^{I,\ell}$ in eq.~\eqref{eq:Pi-I-exact}\,? To facilitate the discussion, we refer the reader to the scheme in fig.~\ref{fig:depletion_scheme}(a), which summarises the backbone of the proposed phenomenology.

%So far we connected the sheet production to the conversion between kinetic and magnetic energies while the sheet saturation, hence the overall sustainment, with the energy transfer across the scales. in particular we saw that both the energy conversion and the energy transfer can be represented by contraction between fields and gradient tensors. We understood that this process is part of the same self-sustaining process as depicted in fig.~\ref{fig:sketch_scheme}, note\footnote{{\red{Probably too Spartan: we might add the curr sheet thinning sketch from the other paper.}}}. The question that arises spontaneously is: can we explain some key features of the energy cascade by taking into account the fact that the strain tends to be aligned with the magnetic field due to the energy conversion? More specifically, can the preferred strain--magnetic-field alignment generated by energy conversion explain why some channels of the cascade are enhanced while others are depleted? there is a scheme\footnote{\red{this scheme has a slightly different philosophy compared to the old (commented out) one. I believe we should add some sketches of VS and SSA with some red crosses. Using at the same time S as a full matrix, and as in diagonal form could be confusing.}}
%It is clear that the prescribed strain magnetic field correlation, namely $\Sbarij \,\overline{b}^\ell_i \, \overline{b}^\ell_j = \varepsilon_b>0$ could a priori reduce the correlation among the strain-rate tensor eigenvalues

More specifically, in HD turbulence we can explain more than $50\%$ of the energy transfer by considering two physical processes: the strain-self amplification and the vortex stretching which are the dominant processes \cite{johnson2020,johnson2021}. The former is the compression and extension of irrotational motion lines while the latter describes the stretching of a vortex tube. Both processes, on average, transfer energy from the large to the small scale \citep{carbone2020,johnson2020,johnson2021}. The gradient expansion introduced above naturally contains these mechanisms: following the notation of eq.~\eqref{eq:notation}, they correspond to $\Pi^{I,\ell}_{SSS}$ and $\Pi^{I,\ell}_{SS\Omega}$, respectively, each comprising single-scale and multiscale contributions.  Specifically, the single-scale contribution associated with strain self-amplification in eq.~\eqref{eq:Pi-I-exact}, denoted in our formalism by $\Pi^{I,\ell}_{s,SSS}$, reads
\begin{equation}
\Pi^{I,\ell}_{s,SSS}
= -\ell^2 \tr{\big(\overline{\vS}^\ell\big)^t
\overline{\vS}^\ell
\big(\overline{\vS}^\ell\big)^t}.
\label{eq:single_SSS}
\end{equation}
Using incompressibility, this can be written entirely in terms of the strain-rate eigenvalues as
\begin{equation}
\Pi^{I,\ell}_{s,SSS}
= -3\, \ell^2 \,\overline{\lambda}^\ell_1 \overline{\lambda}^\ell_2 \overline{\lambda}^\ell_3.
\label{eq:strain_self_single}
\end{equation}
This is precisely proportional to the geometrical quantity introduced earlier to characterise the local strain configuration, a connection long exploited in strain-state analyses \citep{lund1994,johnson2021} and recently recast in an invariant-based description of MHD energy transfer \citep{liptrott2026}.
The depletion of the Inertial transfer suggests a geometrical route through the statistics of the strain. Since strain self-amplification depends on the product $\overline{\lambda}_1^\ell\overline{\lambda}_2^\ell\overline{\lambda}_3^\ell$, its reduction could in principle arise from changes in the magnitude of any of the three eigenvalues. However, while $\overline{\lambda}_1^\ell$ and $\overline{\lambda}_3^\ell$ have fixed signs, only $\overline{\lambda}_2^\ell$ can change sign and therefore produce cancellations in the mean of the overall eigenvalue product. Moreover, the previous section showed that positive energy conversion can be associated with $\overline{\lambda}_2^\ell>0$, providing a second indication that the intermediate eigenvalue might play a pivotal role. We therefore hypothesise that the strain--magnetic-field organisation, from eq.~\eqref{eq:rsc_unf}, weakens this sign preference, making the statistics of $\overline{\lambda}_2^\ell$ more symmetric around zero. Such a change would enhance cancellations between positive and negative events and consequently reduce the mean strain self-amplification. In this sense, the magnetic field would weaken the effective \emph{self-alignment} of the strain compared to the Inertial transfer in HD turbulence.

%without the VS formula
%A complementary picture emerges from vortex stretching, corresponding to $\Pi^{I,\ell}_{S\Omega\Omega}$ in eq.~\eqref{eq:notation}. Although its mean single-scale contribution is related to that of strain self-amplification through the Betchov constraint \cite{betchov1956}, its fluctuations depend on the alignment between vorticity and the strain eigenvectors. As shown by [guys], this preferential alignment is weakened in MHD, while the strain becomes preferentially organised with respect to the magnetic field. This provides a complementary geometrical interpretation of the depletion: the magnetic field reorganises the strain geometry in a way that weakens the vorticity--strain alignment associated with vortex stretching. In addition, \cite{damiano_tsfp_14} recently showed that vortical activity is reduced in MHD compared with HD, indicating that vortex stretching is weakened not only through the altered vorticity--strain alignment, but also through a reduction of the vorticity itself.

A complementary picture emerges from vortex stretching, corresponding to $\Pi^{I,\ell}_{S\Omega\Omega}$ in eq.~\eqref{eq:notation}. Its single-scale contribution can be written as
\begin{equation}
\Pi^{I,\ell}_{s,S\Omega\Omega} = \frac{\ell^2}{4}
|\overline{\bm\omega}^{\ell}|^2 \sum_i \overline{\lambda}_i^\ell \cos^2\theta_{\bm\omega,\ell,i}
\end{equation}
where $\theta_{\bm\omega,\ell,i}$ is the angle between the vorticity and the $i-$th strain eigenvector at scale $\ell$, making explicit its dependence on both the strain eigenvalues and the vorticity--strain alignment \citep{johnson2021}. For the single-scale contributions, its mean is directly related to that of strain self-amplification through the Betchov constraint \cite{betchov1956}, that is $\lan \Pi_{s,SSS}^\ell \ran  = 3 \lan \Pi_{s,S\Omega\Omega}^\ell \ran$, while its fluctuations depend on this geometrical alignment. Recently \cite{liptrott2026} found a strong preferential alignment of vorticity with strain eigenvector relative to $\overline{\lambda}^\ell_2$, that is consistent with the well-known hydrodynamic tendency \citep{ashurst1987,tsinober2001}, and a reduction in the alignment with the eigenvector related to $\overline{\lambda}^\ell_3$. In addition, \cite{damiano_tsfp_14} recently showed that vortical activity is reduced in MHD compared with HD, indicating that vortex stretching is weakened not only through the altered vorticity--strain alignment, but also through a reduction of the vorticity itself. This provides a complementary geometrical interpretation of the depletion: the magnetic field reorganises the strain geometry in a way that weakens the preferential alignment with vorticity responsible for vortex stretching.

The following two sections quantitatively test the two aspects of the proposed phenomenology presented so far: first, the connection between energy conversion and sheet formation, and second, the geometrical mechanism determining the depletion of the Inertial transfer.

\section{Results - sheet production}
\label{sec:results_sheets}

%restore after we move the numerical stuff to the appendix
%In this section we quantify the theoretical picture developed above using data from direct numerical simulations of 3D homogeneous and \emph{isotropic}\footnote{Isotropic in the sense that the equations of motion do not have a preferred direction. See the discussion around the concept of isotropy in MHD in sec.~\ref{sec:sheet_intro}.} and fully developed MHD turbulence at the stationary state in a periodic box that features a unit magnetic Prandtl number $Pm= \nu/\mu  = 1$. Specifically we focus on dataset A2 which is documented in \cite{capocciMHDdata}. 

\purple{In this section we quantify the theoretical picture developed above using data from direct numerical simulations of fully developed, statistically stationary 3D homogeneous MHD turbulence in a periodic box. We deliberately avoid referring to the velocity and magnetic-field fluctuations as \emph{isotropic}: although the equations of motion, eqs.~\eqref{eq:mom_eq}--\eqref{eq:divzeros}, contain no imposed preferred direction, energy conversion generates a local and instantaneous strain--magnetic-field alignment that breaks isotropy, as discussed in sec.~\ref{sec:sheet_intro}. Specifically, we focus on dataset A2 which is documented in \cite{capocciMHDdata}. For the present analysis, we consider the case with no imposed mean magnetic field, which isolates the mechanisms of interest from the additional anisotropy introduced by such a field. We also introduce an auxiliary dataset describing 3D homogeneous and isotropic HD turbulence in the same geometry \citep{buzzicotti2018}; this dataset is considered later in sec.~\ref{sec:results_depletion}. }

\purple{Here, we provide only a brief summary of the numerical simulation setup relevant to the present analysis; key parameters and observables are summarised in table~\ref{tab:datasets}. The MHD data are obtained by solving eqs.~\eqref{eq:mom_eq}--\eqref{eq:divzeros} in a three-dimensional periodic domain using a standard pseudospectral method, whereas the HD dataset is obtained by solving the incompressible Navier--Stokes equations, i.e. the equations resulting from setting the magnetic field equal to zero, in the same geometry and within the same numerical framework. Time integration is performed using a second-order Runge--Kutta scheme, with dealiasing implemented through the two-thirds rule \citep{patterson1971,canuto1988}. The MHD dataset has unit magnetic Prandtl number, $Pm=\nu/\mu=1$. In both datasets, the mechanical forcing in eq.~\eqref{eq:mom_eq} is a drag-free Ornstein--Uhlenbeck process, active in the wavenumber band $k\in[2.5,5.0]$ for MHD and $k\in[0.5,1.5]$ for HD. }
\begin{table}
  \begin{center}
\def~{\hphantom{0}}
   \begin{tabular}{ccccccccccccccccc}
        \hline
        \hline
		 id & $N$ & $\langle E_u \rangle$ & $\langle E_b \rangle$ & $\nu/10^{-4}$ & $\eps_u$ & $\eps_b$ &  $\mbox{Re}$ &  $L_f/\eta$ & $k_{max}\, \eta$ \\
        \hline %tau = 0.79 (MHD) tau = 0.83 (HD)
      A2 & 2048 & 0.73 & 0.38 & $2$ & 0.22  & 0.52  & 2144  & 467 & 1.2    \\
      \hline
      V2 & 2048 & 1.12 & $-$ & $3$ & 1.4  & $-$  & 7000  & 1495 & 1.4    \\
      \hline
      \hline
        \end{tabular} 
        \caption{
\purple{Simulation parameters and key observables, where
        $N$ is the number of collocation points in each coordinate,
        $\lan E_u \ran $ and $\lan E_b \ran $ are the mean kinetic and magnetic energies respectively,
        $\nu$ the kinematic viscosity,
        $\eps_u$ and $\eps_b$ are the kinetic and magnetic energy dissipation rates,
        and $\mbox{Re}$ is the Reynolds number. 
	  In addition, $L_f/\eta$ is the ratio between the forcing scale and the Kolmogorov scale $\eta= (\nu^3 / \varepsilon)^{1 / 4}$, and 
        $k_\text{max}$ the largest retained wavenumber component after de-aliasing which is multiplied by the Kolmogorov scale. 
        The magnetic Prandtl number, 
        $Pm = \nu / \mu  $, which is the ratio between the viscosity and magnetic diffusivity, equals unity for A2 while V2 is a dataset describing hydrodynamics.}
        }
  \label{tab:datasets}
  \end{center}
\end{table}

%The central question is how the flow organises to realise the net positive kinetic-to-magnetic energy conversion imposed by the stationary energy balance: which strain geometries and strain--magnetic-field alignments are preferentially selected and, in particular, is this conversion predominantly associated with sheet-like or tubular velocity structures? To answer this, we return to the eigenvalue representation of the resolved-scale energy conversion introduced in eq.~\eqref{eq:conv_spec} that we straightforwardly generalise to the resolved-scale energy conversion of eq.~\eqref{eq:RSC}
\purple{With the numerical setup specified, we now turn to the central question: how does the flow organise to realise the net positive kinetic-to-magnetic energy conversion imposed by the statistically stationary energy balance?} Which strain geometries and strain--magnetic-field alignments are preferentially selected and, in particular, is this conversion predominantly associated with sheet-like or tubular velocity structures? To answer this, we return to the eigenvalue representation of the resolved-scale energy conversion introduced in eq.~\eqref{eq:conv_spec} that we straightforwardly generalise to the resolved-scale energy conversion of eq.~\eqref{eq:RSC}
\begin{equation}
    \mathcal{W}^\ell =  |\overline{\bm{b}}^\ell|^2   \overline{\lambda}^\ell_i \cos^2{\theta_{\bm{b},\ell,i}}
    \label{eq:conv_spec_filter}
\end{equation}
By applying filtering method, we can assess how the relationship between fields geometry and energy conversion changes at different length-scales. In principle, as discussed above, a complete statistical characterisation requires considering the joint statistics of $\mathcal{W}^\ell$ with each variable entering eq.~\eqref{eq:conv_spec_filter}, as well as with their combined contribution. In this section, we focus on $P({\mathcal{W}}^\ell,\overline{\lambda}^\ell_2)$, as $\overline{\lambda}^\ell_2$ is the eigenvalue sign-indefinite. Specifically, as introduced in sec.~\ref{sec:sheet_intro}, if positive energy conversion,  $\mathcal{W}^\ell>0$, occurs preferentially when $\overline{\lambda}^\ell_2>0$, then kinetic-to-magnetic energy conversion can be associated with sheet-like structures formation. To assess the length scale dependency of this association, we consider both the unfiltered case, $\ell=0$, and $\ell\approx 59 \, \eta$ corresponding to the inertial range.
\begin{figure}
	\centering
    \noindent\makebox[\textwidth]{
    \includegraphics[width=0.58\columnwidth]{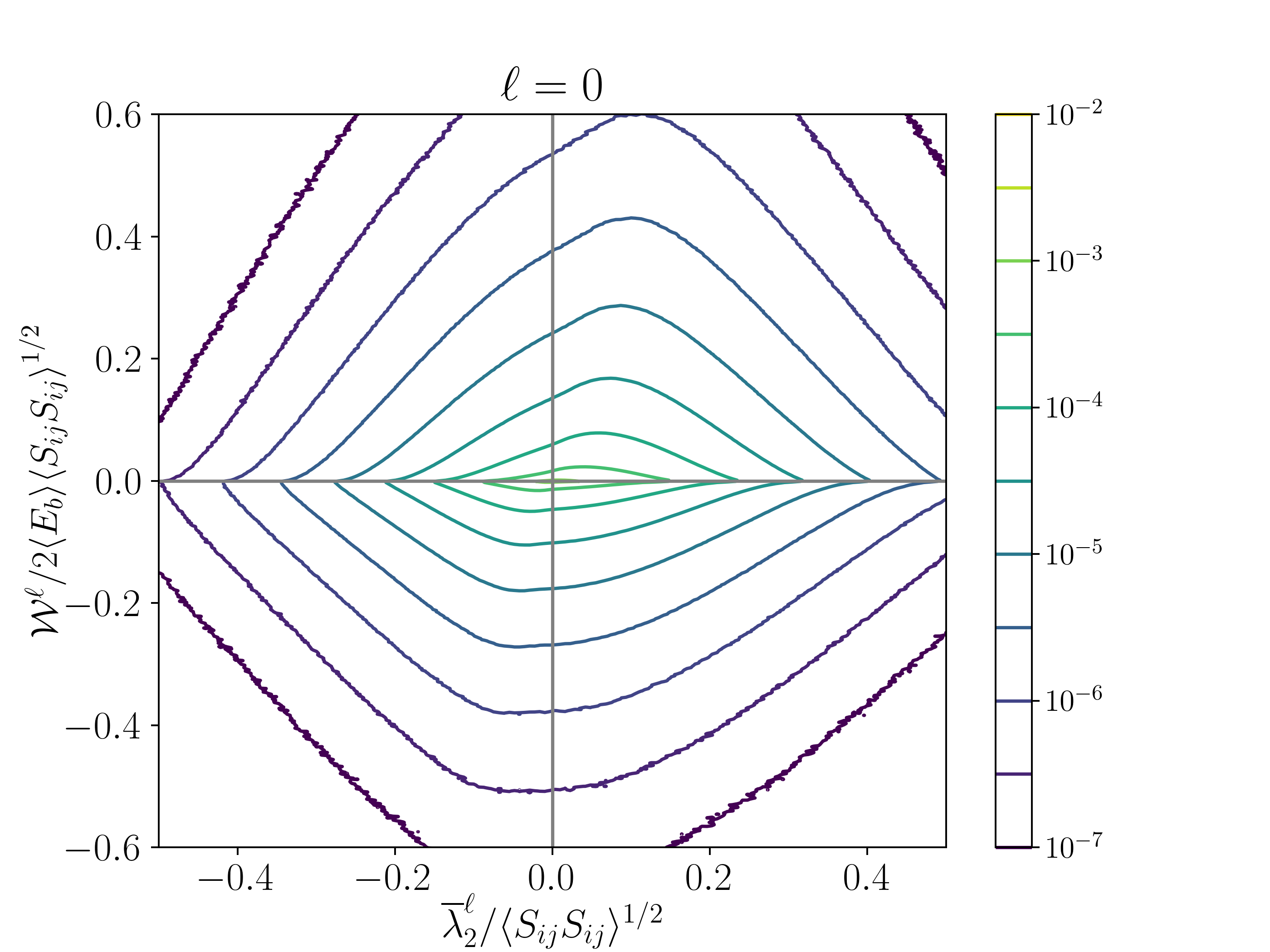} 
    \hspace{-1.5cm}
    \includegraphics[width=0.58\columnwidth]{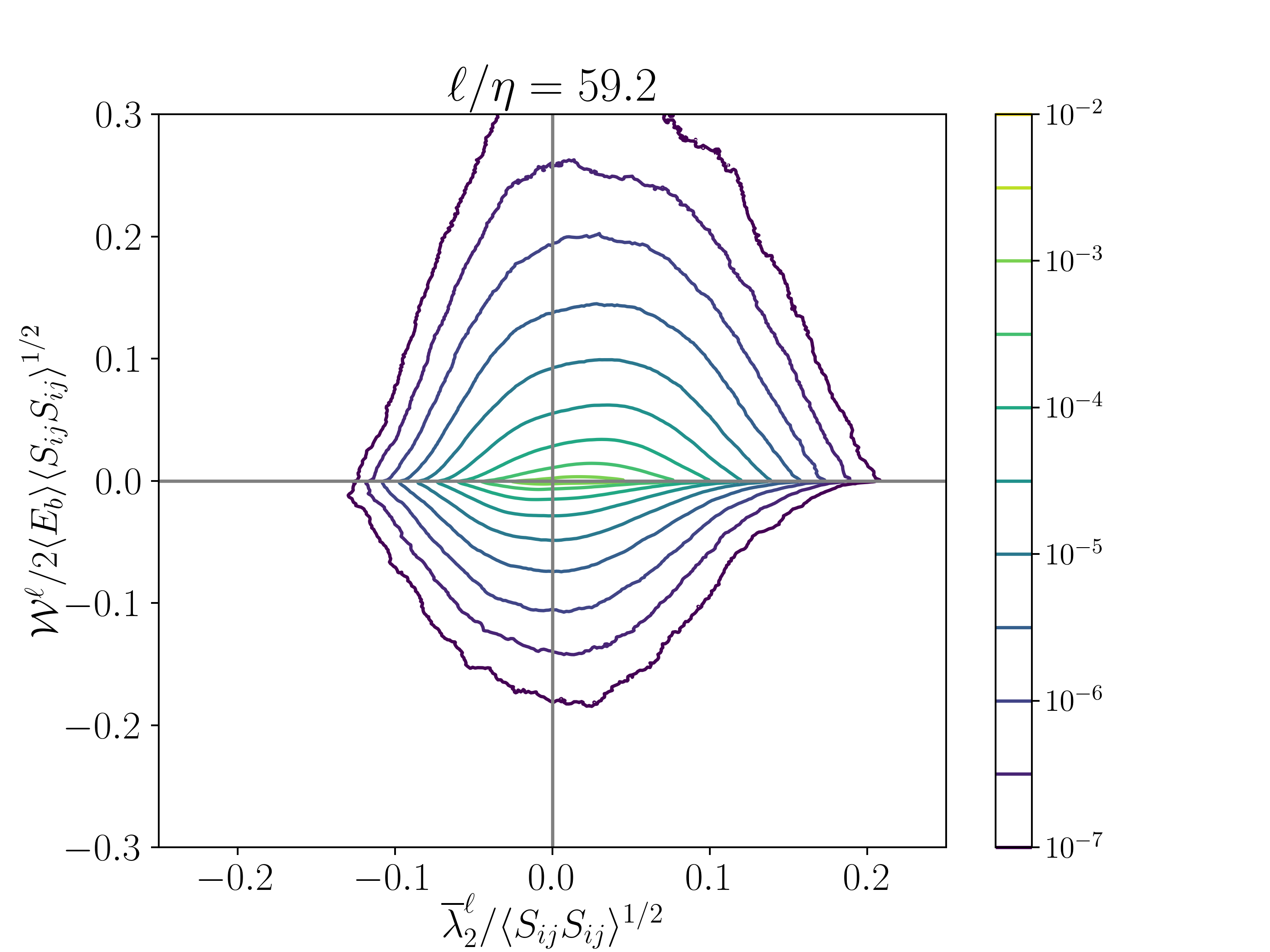} 
    }
	 \caption{
Isolines of the joint pdf between $\mathcal{W}^\ell$ and the intermediate strain-rate tensor eigenvalue $\overline{\lambda}_2^\ell$, related to the full field (left panel) and to the Inertial range (right panel). The aspect ratio of the right panel is the same as that of the left panel, although the axis ranges are different.}
\label{fig:joint_W_second_eig}
\end{figure}

%\marginnote{\textbf{Joint pdf with the intermediate eigenvalue}}
%In fig.~\ref{fig:joint_W_second_eig}, we observe that events with $\mathcal{W}^\ell>0$ and $\overline{\lambda}^\ell_2>0$ are more likely to happen than $\mathcal{W}^\ell>0$ and $\overline{\lambda}^\ell_2<0$ since the joint pdfs are more outbalanced in favour of the former. This indicates that the kinetic-to-magnetic energy conversion is realised by sheet-like structures. 
In fig.~\ref{fig:joint_W_second_eig}, the positive-conversion branch of the isolines is clearly biased towards $\overline{\lambda}_2^\ell>0$, with the strongest positive-conversion events occurring preferentially for positive intermediate eigenvalues. Since $\overline{\lambda}_2^\ell>0$ corresponds to two extensional and one compressive strain directions, strong kinetic-to-magnetic energy conversion is therefore preferentially associated with sheet-like velocity structures. Note that the isolines are not $y-$axis symmetric in consequence of the energy balance, i.e. $\lan \mathcal{W}^\ell \ran > 0 $. On the contrary, in the third and fourth quadrants i.e. $P(\mathcal{W}^\ell<0,  \overline{\lambda}^\ell_2 )$, the isolines have a more pronounced $x-$axis symmetry meaning that sheets and tubular/filamentous structures contribute equally to the magnetic-to-kinetic energy transfer, viz $\mathcal{W}^\ell<0$. This applies to both scales considered here, corresponding to the two panels in fig.~\ref{fig:joint_W_second_eig}, as well as to the scales within them (not shown). As regards the joint statistics between $\mathcal{W}^\ell$ and the remaining degrees of freedom entering the energy conversion, we briefly summarise their main features, while a more detailed quantification is provided in Appendix~\ref{sec:appendix_1}. 

For the remaining strain eigenvalues, $P\left(\mathcal{W}^\ell,\overline{\lambda}^\ell_1\right)$ and $P\left(\mathcal{W}^\ell,\overline{\lambda}^\ell_3\right)$ show no clear association between regions of  large strain and strong energy-conversion events.
%\marginnote{\textbf{Joint pdf with the other freedoms}}
Concerning the magnetic-field orientation relative to the strain motion, we consider $P(\mathcal{W}^\ell,\theta_{\bm{b},\ell,i})$. The magnetic field-strain alignment associated with strong positive conversion displays a clear scale dependence. For the unfiltered fields, large energy conversion is associated with strong alignment with the eigenvector associated with $\overline{\lambda}_2^\ell$, with $\theta_{\bm{b},\ell,2}\simeq0^\circ$. Conversely, events of intense and positive energy conversion occur in presence of more oblique orientations, of approximately $42^\circ$, for the eigenvectors associated with the two sign-definite eigenvalues, i.e. $\overline{\lambda}_1^\ell$ and $\overline{\lambda}_3^\ell$. As the filtering scale increases, strong positive conversion remains preferentially associated with alignment along the eigenvector of $\overline{\lambda}_2^\ell$, although with a broader angular selectivity. At the same time, strong positive conversion becomes associated with magnetic-field orthogonality to the eigenvector of the compressive eigenvalue, i.e. $\theta_{\bm{b},\ell,1}\simeq90^\circ$, and alignment with that of the extensional eigenvalue, viz $\theta_{\bm{b},\ell,3}=0^\circ$.

Finally, from $P(\mathcal{W}^\ell, |\overline{\bm{b}}^\ell|^2 )$, we do not observe a clear association between large values of energy conversion and regions of large magnetic energy.

\purple{The analysis presented here, together with the joint statistics involving the remaining strain eigenvalues, the strain--magnetic-field angles, the magnetic-field strength and the combined strain--alignment contribution discussed in Appendix~\ref{sec:appendix_1}, was repeated on a dataset at lower resolution and Reynolds number to verify the convergence of the observed trends.}
%I used "was" as the subject is still "the analysis presented here". I never know how to prceed with these things.

Taken together, these results support the theoretical picture developed in sec.~\ref{sec:sheet_intro}: positive kinetic-to-magnetic energy conversion is preferentially associated with sheet-like strain geometries across scales, while the detailed strain--magnetic-field organisation varies with scale and results from the combined interplay of field strength, strain intensity and magnetic field-strain motion alignment. Having established this first part of the phenomenology, we now ask whether the same reorganisation of the strain field can account for the depletion of the Inertial transfer, as proposed in sec.~\ref{sec:depl_inertial_theory}.

%note that tubes at small scale might be embedded withing sheets but sheets are always overabundant over filaments}

\begin{comment}
\vspace{2cm}
\textbf{check the values of the normalisation - just to help the writing}\\
\begin{tabular}{|c|c|c|}
     \hline
     \multicolumn{3}{|c|}{maxima of $\mathcal{W}^\ell = |\overline{\bm{b}}^\ell|^2   \overline{\lambda}^\ell_i \cos^2{\theta_{\overline{\bm{b}},i}}$}\\
     \hline
     observable  & $\ell \approx 0$  & $\ell \in$ inertial range  \\
     \hline
     \hline
     $\theta_1^\ell $ &    $[0,\, \theta^{*}=42^\circ]$  &  $90^\circ$  \\
          \hline
     $\theta_2^\ell$  &    0                          &  $ [0,  \, 2\pi]$ \\     
          \hline
     $\theta_3^\ell$  &     $ \gtrsim \theta^*$     &  0  \\
          \hline
     $-|\overline{\lambda}_1^\ell|$   &    $\approx-0.50\,\lan {S}_{ij} {S}_{ij} \ran^{1/2}$      &  $\approx-0.12\,\lan {S}_{ij} {S}_{ij} \ran^{1/2}$  \\
          \hline
     $\overline{\lambda}_2^\ell$   &    $\approx 0.08\,\lan {S}_{ij} {S}_{ij} \ran^{1/2}$      &  $\approx 0.03\,\lan {S}_{ij} {S}_{ij} \ran^{1/2}$  \\
     \hline
     $\overline{\lambda}_3^\ell$   &  $\approx 0.35\,\lan {S}_{ij} {S}_{ij} \ran^{1/2}$      &  $\approx 0.12\,\lan {S}_{ij} {S}_{ij} \ran^{1/2}$  \\
     \hline
     $E_b^\ell$   &  \emph{intermediate}     &  upper \emph{intermediate}  \\
     \hline
     $ \overline{\lambda}^\ell_i \cos^2{\theta_{\overline{\bm{b}},i}}$   &  $7.0\times 10^{-3}/2 \lan {S}_{ij} {S}_{ij} \ran $     &  $6.8\times 10^{-3} \lan {S}_{ij} {S}_{ij} \ran  $  \\
     \hline
\end{tabular}
\end{comment}

\section{Results - Depletion of Inertial transfer}
\label{sec:results_depletion}

%\marginnote{\textbf{Show the Inertial trans depletion}}

We next test the prediction developed in sec.~\ref{sec:depl_inertial_theory}, namely that the strain reorganisation associated with energy conversion underlies the depletion of the Inertial transfer in MHD relative to HD. As such, we first quantify the depletion itself. 
%To calculate the fluxes in HD, we consider an auxiliary dataset of fully developed homogeneous and isotropic turbulence at the stationary state, corresponding to the database V2 from \cite{buzzicotti2018}.
To calculate the fluxes in HD, we consider \purple{the auxiliary dataset V2 introduced in sec.~\ref{sec:results_sheets}}. \purple{We stress that the depletion of the mean Inertial transfer has also been observed in simulations with hyperviscous dissipation \citep{capocci2025} and with a (strong) mean magnetic field at lower resolution \citep{capocci2026}. This robustness across different physical and numerical settings motivates our focus on a single dataset in the present analysis.}

\begin{figure}
	\centering
    %\noindent\makebox[\textwidth]{
    \includegraphics[width=0.9\columnwidth]{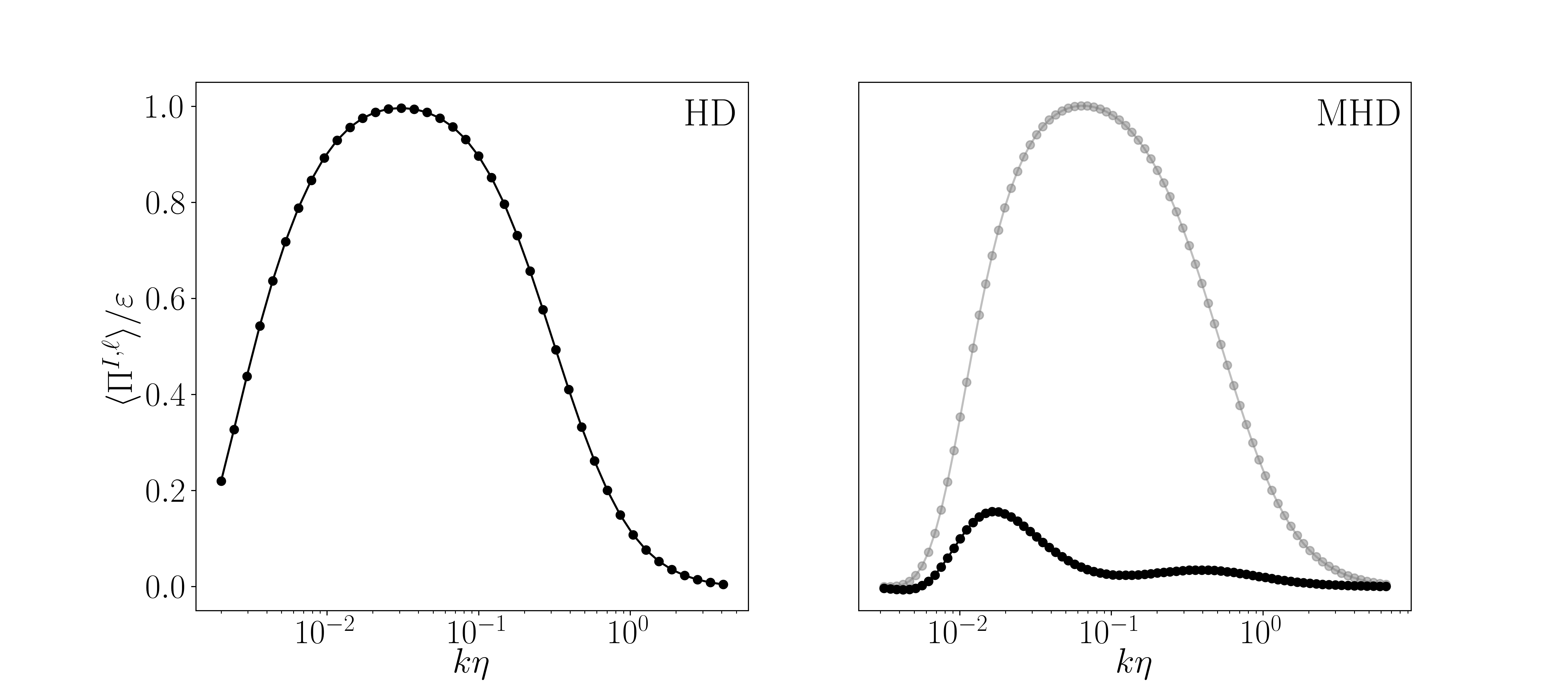} 
    %}
	 \caption{Inertial transfer as a function of the non-dimensional  length-scale $k\eta = \pi \, \eta /\ell$ and normalised by the total energy dissipation rate $\varepsilon$ for HD (left panel) and MHD (right panel). The black curves correspond to the Inertial transfer term, eq.~\eqref{eq:Pi-I-exact} while the gray shaded curve describes the total energy transfer in MHD, namely the sum of eqs.~\eqref{eq:Pi-I-exact}--~\eqref{eq:Pi-D-exact}. Both axes show non-dimensional quantities. Note that the value $k \eta = 0.05$ corresponds to the scale associated with the right panel of fig.~\ref{fig:joint_W_second_eig}. }
\label{fig:total_fluxes}
\end{figure}

Fig.~\ref{fig:total_fluxes} compares the Inertial transfer, $\Pi^{I,\ell}$, defined by the left-hand side of eq.~\eqref{eq:Pi-I-exact}, in MHD and HD. In MHD, $\Pi^{I,\ell}$ is only one component of the total interscale energy transfer, whereas in HD it constitutes the full interscale flux. The comparison clearly indicates that the Inertial contribution is strongly reduced in MHD across scales accounting for less than $5\%$ at the total energy transfer peak, which is displayed by the grey curve.

%\marginnote{\textbf{Etiology of depletion - $\lambda_2$}}

\begin{figure}
	\centering
    %\noindent\makebox[\textwidth]{
    \includegraphics[width=0.9\columnwidth]{ 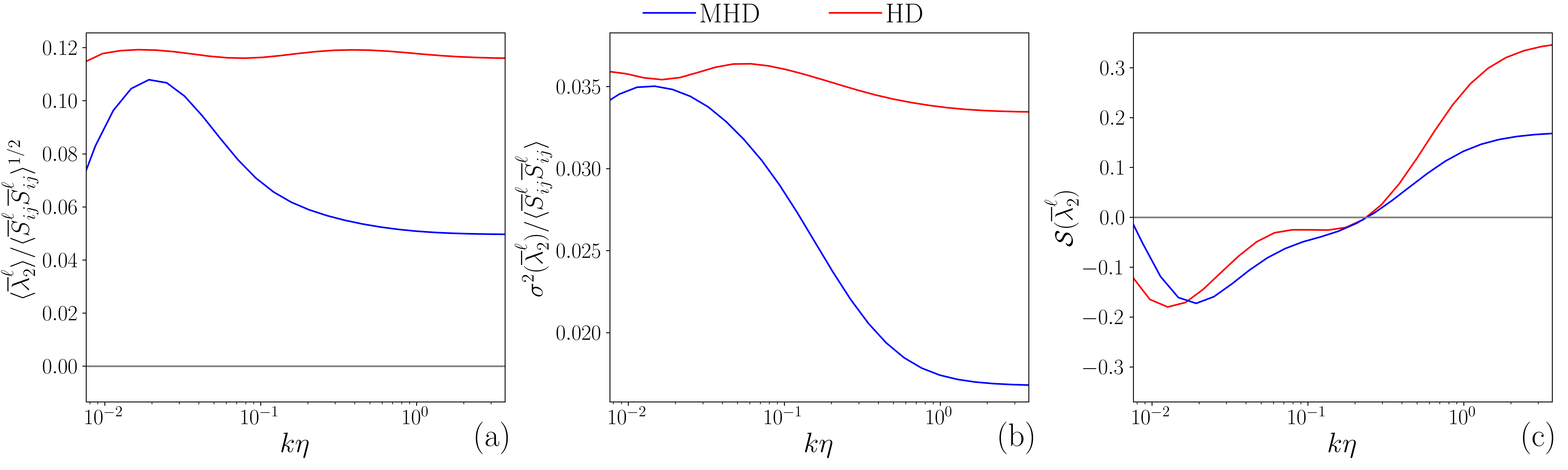} 
    %}
	 \caption{Mean, variance and skewness of the intermediate strain rate tensor eigenvalue $\overline{\lambda}^\ell_2$ as functions of the non-dimensionalised scale $k\eta = \pi \,\eta/\ell$. Both mean and standard deviation are made non-dimensional by powers of the mean strain rate $\lan S_{ij}\, S_{ij} \ran$.}
\label{fig:second_eig_moments}
\end{figure}

We now examine the geometrical origin of this depletion, first through the statistics of $\overline{\lambda}_2^\ell$ and then through the eigenvalue product $\overline{\lambda}_1^\ell\overline{\lambda}_2^\ell\overline{\lambda}_3^\ell$ entering eq.~\eqref{eq:strain_self_single}. Fig.~\ref{fig:second_eig_moments} compares mean, variance and skewness of $\overline{\lambda}_2^\ell$ across the scales between HD and MHD, with the first two central moments normalised by the appropriate power of the averaged strain-rate norm. In MHD, $\overline{\lambda}_2^\ell$ is generally smaller on average, and displays weaker fluctuations and is less skewed than in HD, with these differences becoming increasingly pronounced towards smaller scales. These results are consistent with the theoretical picture developed in sec.~\ref{sec:depl_inertial_theory}: the bias of the intermediate eigenvalue towards positive values is weakened in MHD, making its statistics more symmetric and thereby favouring stronger cancellations in the mean of strain self-amplification of eq.~\eqref{eq:single_SSS}. The reduction is particularly significant because the mean strain-rate squared, $\lan S_{ij}\, S_{ij} \ran $ used for normalisation, is itself smaller in MHD than in HD. The normalisation therefore acts against, rather than produces, the observed decrease in the mean and fluctuations of $\overline{\lambda}_2^\ell$. 

\begin{figure}
	\centering
    %\noindent\makebox[\textwidth]{
    \includegraphics[width=0.9\columnwidth]{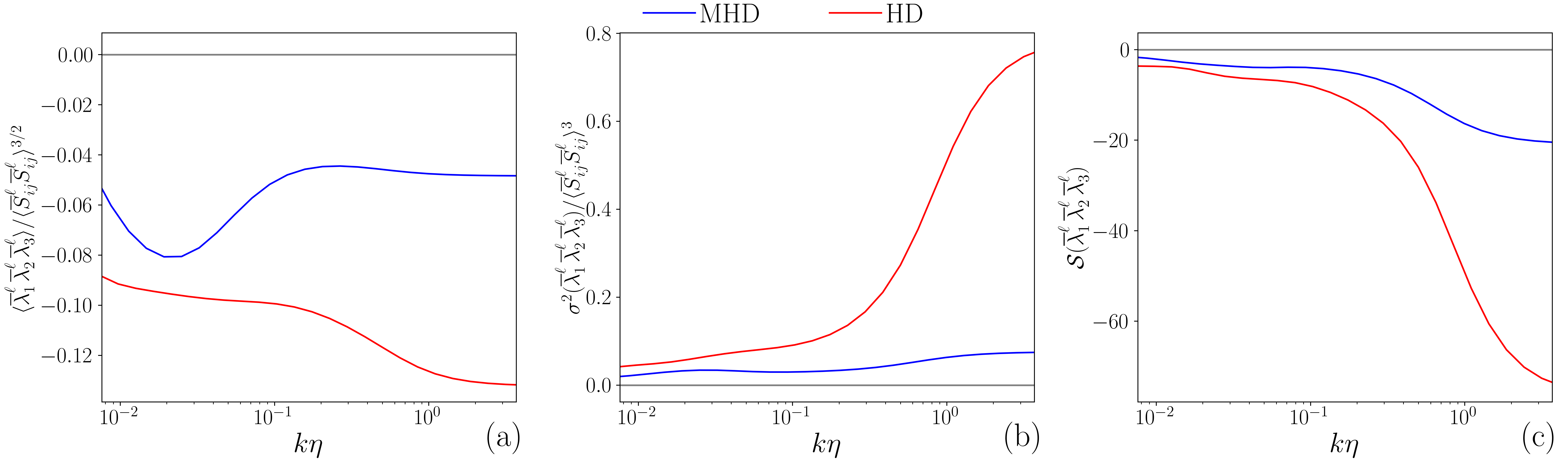} 
    %}
	 \caption{Mean, variance and skewness of the strain-rate tensor eigenvalue product. $\overline{\lambda}_1^\ell\overline{\lambda}_2^\ell\overline{\lambda}_3^\ell$ as functions of the scale $k\eta = \pi \,\eta/\ell$. Both mean and standard deviation are made non-dimensional by powers of the mean strain rate $\lan S_{ij}\, S_{ij} \ran$.}
\label{fig:second_eig_prod_moments}
\end{figure}

%\marginnote{\textbf{Etiology of the depletion - $\lambda_1 {\lambda}_2 {\lambda}_3$}}
As anticipated above, we now turn to the full strain-rate tensor eigenvalue product $\overline{\lambda}_1^\ell\overline{\lambda}_2^\ell\overline{\lambda}_3^\ell$.
Similarly to fig.~\ref{fig:second_eig_moments}, fig.~\ref{fig:second_eig_prod_moments} displays the corresponding first three central moments as a function of the filtering scale. This analysis can be viewed either as an energy-transfer discussion, following sec.~\ref{sec:depl_inertial_theory}, or as a flow geometry study, in the spirit of sec.~\ref{sec:sheet_intro}. This follows from the interpretation of the product of the strain-rate tensor eigenvalues both as a classifier of the flow geometry and as a quantity entering the energy transfer across scales. In fact, the first feature we observe in fig.~\ref{fig:second_eig_prod_moments}(a) is that, at small scales, i.e. for large $k\eta = \pi \eta/\ell$, MHD exhibits less intense sheet-like structures than HD, as predicted by eq.~\eqref{eq:eig_prod_eq} and discussed in sec.~\ref{sec:sheet_intro}. Conversely, when interpreted through its energy-transfer role, the reduction of the eigenvalue product directly reflects the depletion of strain self-amplification, and hence of the Inertial transfer, in MHD relative to HD. From intermediate to small scales, this reduction in the mean is accompanied by weaker fluctuations and reduced skewness, as shown in fig.~\ref{fig:second_eig_prod_moments}~(b)-(c), respectively. The distribution $P(\overline{\lambda}_1^\ell\overline{\lambda}_2^\ell\overline{\lambda}_3^\ell)$ therefore becomes more balanced between positive and negative events, leading to stronger cancellations in its mean. Like in fig.~\ref{fig:second_eig_moments}, the adopted normalisation, in both mean and standard deviation, makes these differences even more significant. These results therefore confirm the prediction of sec.~\ref{sec:depl_inertial_theory}: relative to HD, MHD exhibits a reorganisation of the strain statistics that enhances cancellations in the \purple{mean} eigenvalue product and suppresses strain self-amplification and the associated Inertial transfer.

\section{Experimental observations and predictions }
\label{sec:experiments}

Having discussed the theoretical predictions and methodology and their confirmation through numerical simulations, we now focus on experimental realisations. Specifically, we structure this section around two questions: i) which quantities does our analysis suggest should be measured experimentally? ii) which existing measurements already provide support for the picture developed above?

These questions should be addressed in the most natural experimental context which is the spacecraft observations where satellites allow for in situ measurements of astrophysical plasma fields, see \cite{klein2023} which will be discussed later. In addition, the coarse graining operation, that plays a crucial role in the theory developed herein, can be approximated experimentally through local multi-spacecraft measurements at finite spatial separations. Such a coarse graining is not, however, realised by the low-pass filter as in eq.~\eqref{eq:coarse_gr} but via the network of satellites that form tetrahedron method, see \cite{consolini2015} and references therein. The limitations imposed by the finite number of simultaneously accessible spatial scales, and the prospects offered by future multi-spacecraft missions, are discussed below.

The geometrical quantities required by the present theory are therefore experimentally accessible. In particular, coarse-grained velocity- and magnetic-field gradient tensors, together with their invariants and associated topologies, have already been reconstructed from spacecraft measurements in turbulent space plasmas \citep{consolini2015,quattrociocchi2019,quattrociocchi2025,hnat2021}. Importantly, the application of the gradient-based flux decomposition introduced in \cite{capocci2025} to multi-spacecraft measurements has recently been proposed explicitly by \cite{liptrott2026}, through an invariant-based formulation in which experimentally accessible gradient tensors provide proxies and bounds for the local mechanistic energy fluxes. 

%in quattrociocchi2025, they show some interesting ratios however the difference in the homogeneisation of fields gradients could lead to wrong results.

An observational connection between magnetic-field topology and interscale energy transfer has recently been established by \cite{hnat2025}, who found that forward transfer is preferentially associated with hyperbolic, three-dimensional X-line magnetic topology. Although velocity increments enter their energy transfer estimate, 
%the velocity gradient geometry associated with these magnetic structures is not characterised; 
\purple{the velocity-gradient geometry associated with these magnetic structures was not the focus of that analysis;}
the single-scale gradient decomposition adopted here could therefore complement such measurements by identifying the specific local deformation mechanisms contributing to the observed transfer.

A particularly relevant experimental counterpart to the present analysis is provided by the recent spacecraft observations of turbulent dynamo activity in \cite{voros2026}. Their results allow us to identify both aspects of the present picture that already find observational support and quantities that could be measured to test its more specific predictions. Specifically, \cite{voros2026} analysed spacecraft data to quantify the local contributions governing magnetic-field amplification and suppression, including the field-aligned velocity gradient term and compressive effects. Their observations therefore provide direct experimental evidence for the local realisation of velocity gradient-driven magnetic amplification in a turbulent plasma. Moreover, the \emph{stretched} and \emph{folded} magnetic configurations identified in their analysis are consistent with the development of strong magnetic gradients and current-sheet-like structures. 
\purple{Their energy-conversion analysis focuses on the scalar quantity $b_i b_j S_{ij}/b^2$, which is similar to eq.~\eqref{eq:rsc_unf}, rather than resolving separately the strain geometries and magnetic-field alignments through which the same positive conversion may be realised.}
%Their energy conversion analysis, however, treats $ b_i  b_j S_{ij}/b^2$, which is similar to eq.~\eqref{eq:rsc_unf}, as a single scalar quantity and does not resolve the different strain geometries and magnetic-field alignments through which the same positive conversion may be realised. 
In addition, their stretch-and-fold picture captures the kinematic action of velocity gradients on the magnetic field: \cite{voros2026} explicitly adopt a kinematic-dynamo formulation in which magnetic back-reaction on the plasma is neglected. 
\purple{The reciprocal modification of the velocity-strain geometry by the magnetic structures generated through this process therefore lies outside the scope of their analysis.}
%Their analysis therefore does not address the reciprocal modification of the velocity-strain geometry by the magnetic structures generated through this process. 
In the framework proposed here, this constitutes \purple{one} branch of a coupled dynamical loop: energy conversion organises the velocity-strain geometry, the resulting deformation generates and sharpens magnetic gradients, and current-sheet thinning contributes to the interscale transfer while feeding back on the velocity field. The present analysis also suggests concrete measurements through which the present picture could be tested. Since the conversion term can be decomposed as
eq.~\eqref{eq:conv_spec_filter}, 
multi-spacecraft measurements of the velocity gradient tensor could decompose the observed stretching rate into its individual geometrical degrees of freedom, as done in eqs.~\eqref{eq:conv_spec} and \eqref{eq:conv_spec_filter}. In particular, joint statistics between energy conversion and the strain-rate tensor eigenvalues and the corresponding magnetic-field alignments would then determine which local deformation geometries preferentially realise positive kinetic-to-magnetic conversion. In the compressible environments considered by \cite{voros2026}, the isotropic part of the velocity gradient could be separated from its deviatoric strain, allowing the corresponding strain invariants and eigenvalues to test whether the geometrical picture identified here remains robust when compressibility is present. These geometrical measurements could then be directly connected to the specific mechanistic contributions to interscale energy transfer identified by the gradient decomposition discussed above.

A present \purple{experimental limitation} is that a four-spacecraft tetrahedral configuration provides essentially a single characteristic three-dimensional sampling scale for direct gradient estimates at a given time. Future satellites networks such as HelioSwarm \cite{klein2023}, with nine spacecraft and up to 126 possible tetrahedral configurations over separations of approximately $50$--$3000 \,\mathrm{km}$, could overcome this limitation. This would allow the scale-dependent geometrical and energetic mechanisms discussed here to be tested directly across multiple spatial scales.

\section{Conclusions}

In this work, we investigated how the energy exchange in magnetohydrodynamic flows can impose a preferred flow geometry. We showed that the formation of current sheets in magnetohydrodynamics arises naturally from the exchange of energy between the kinetic and magnetic channels. This connection is further supported by analytical arguments, which indicate that in MHD sheet-like velocity geometries constitute the dominant configuration against filament-like structures. We further showed that sheet-like structures in MHD are more anisotropic than their HD counterparts, exhibiting weaker extension along the eigenvector associated with the intermediate strain eigenvalue.

Moreover, the coupling between energy conversion and flow geometry provides a physical explanation for the depletion of the Inertial energy transfer in MHD relative to HD. This effect is mainly associated with the enhanced symmetry of the probability density function of the intermediate eigenvalue of the strain-rate tensor which increases cancellations between positive and negative values in the mean strain self-amplification. This explains the previously observed depletion of Inertial transfer in MHD in terms of the energy conversion-driven reorganisation of the velocity gradient geometry. 

Recent spacecraft observations already provide experimental support for parts of this picture, in particular the role of velocity-gradient deformation in magnetic-field amplification and the associated development of structured magnetic configurations. At the same time, the present framework identifies more specific observables through which the proposed mechanism can be tested, including strain-eigenvalue statistics, strain--magnetic-field alignments, gradient-based contributions to interscale energy transfer, and their scale dependence. Future multi-spacecraft \emph{constellations} with simultaneous access to multiple spatial scales may therefore provide a direct experimental test of the coupled geometrical and interscale energy transfer loop proposed here.

\clearpage

\appendix

\section{Additional joint statistics}
\label{sec:appendix_1}

\begin{figure}
	\centering
    \noindent\makebox[\textwidth]{
    \includegraphics[width=0.5\columnwidth]{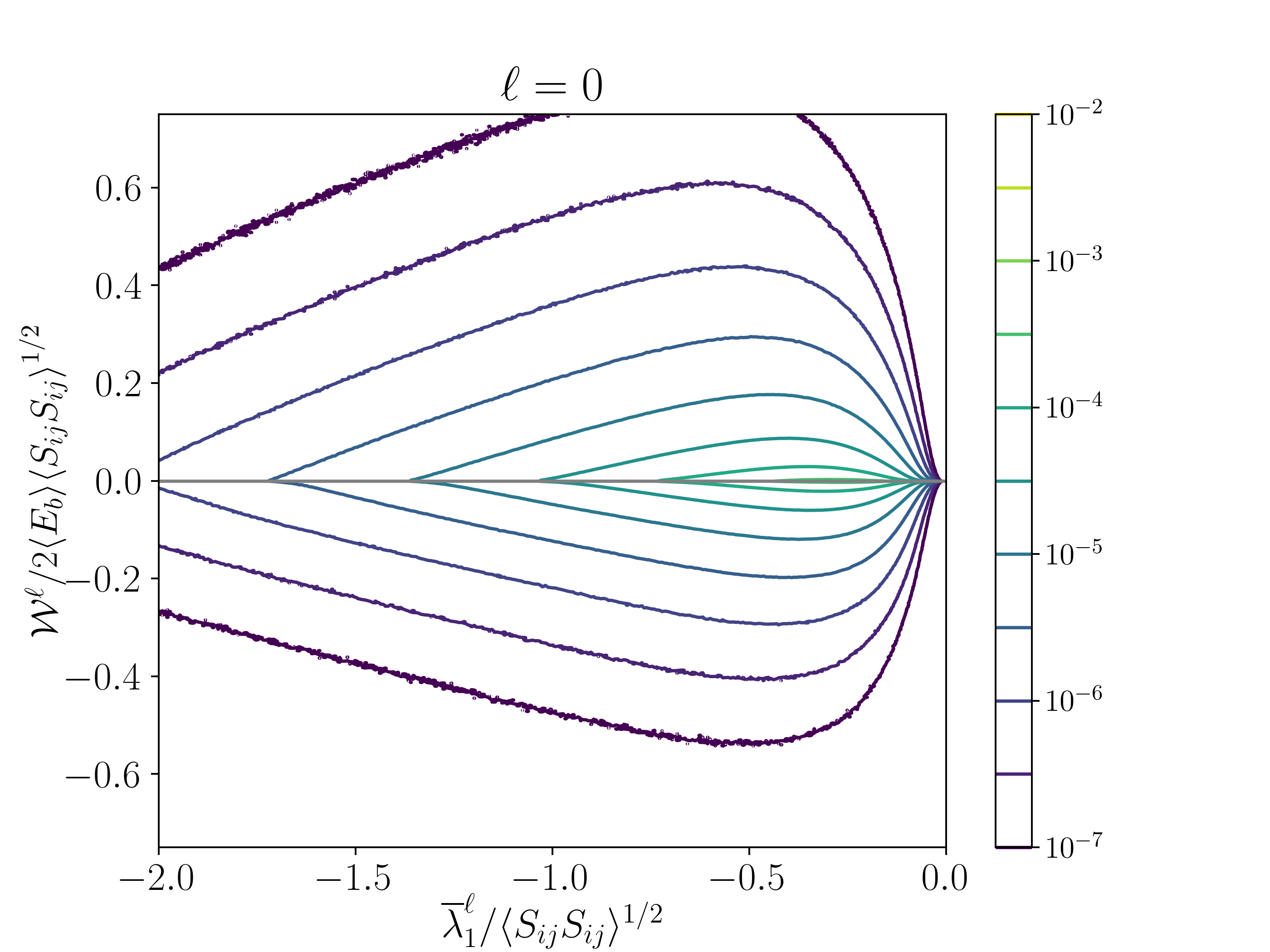} 
    \hspace{-1.5cm}
    \includegraphics[width=0.5\columnwidth]{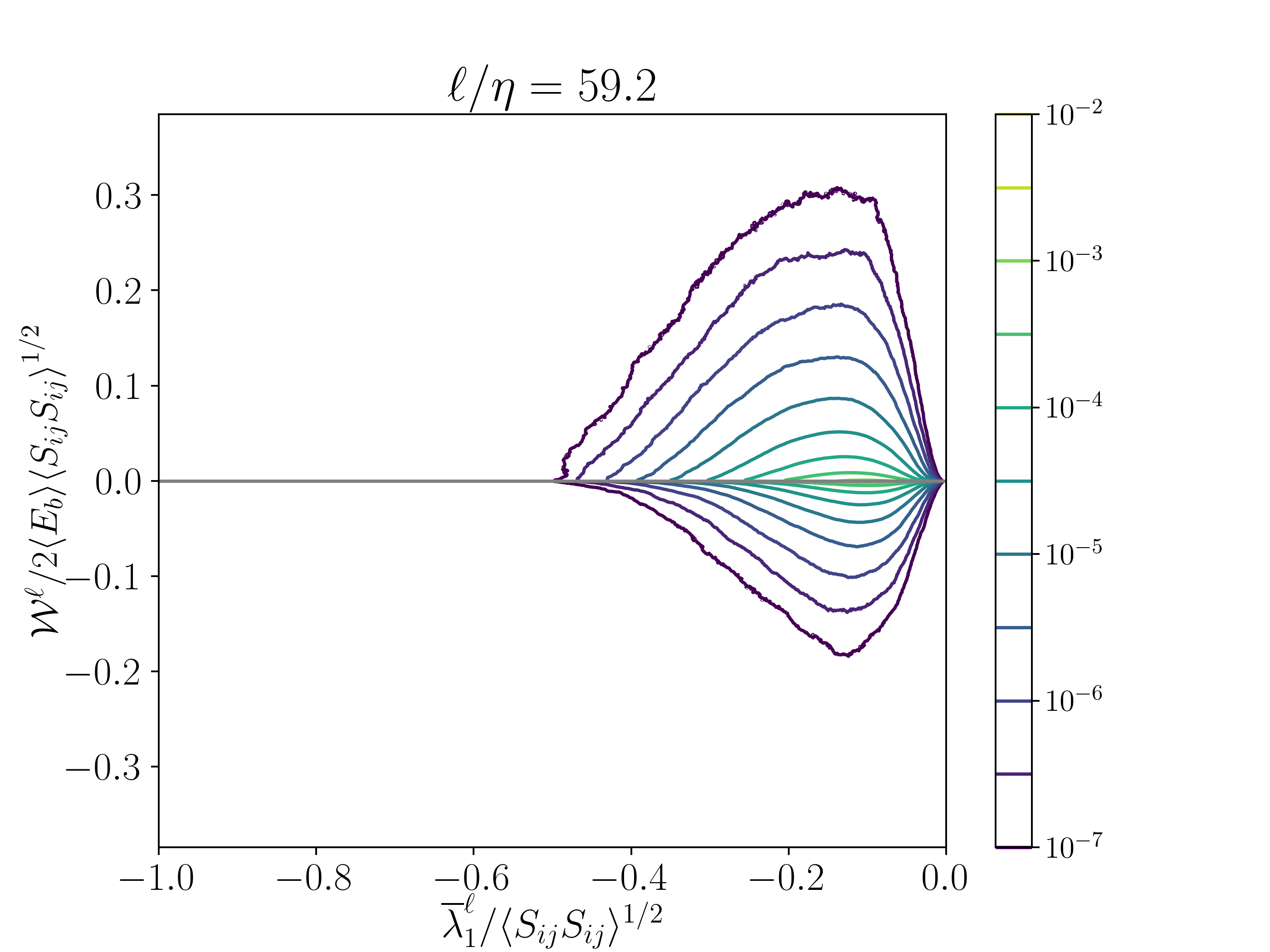}
    }
    
    \includegraphics[width=0.5\columnwidth]{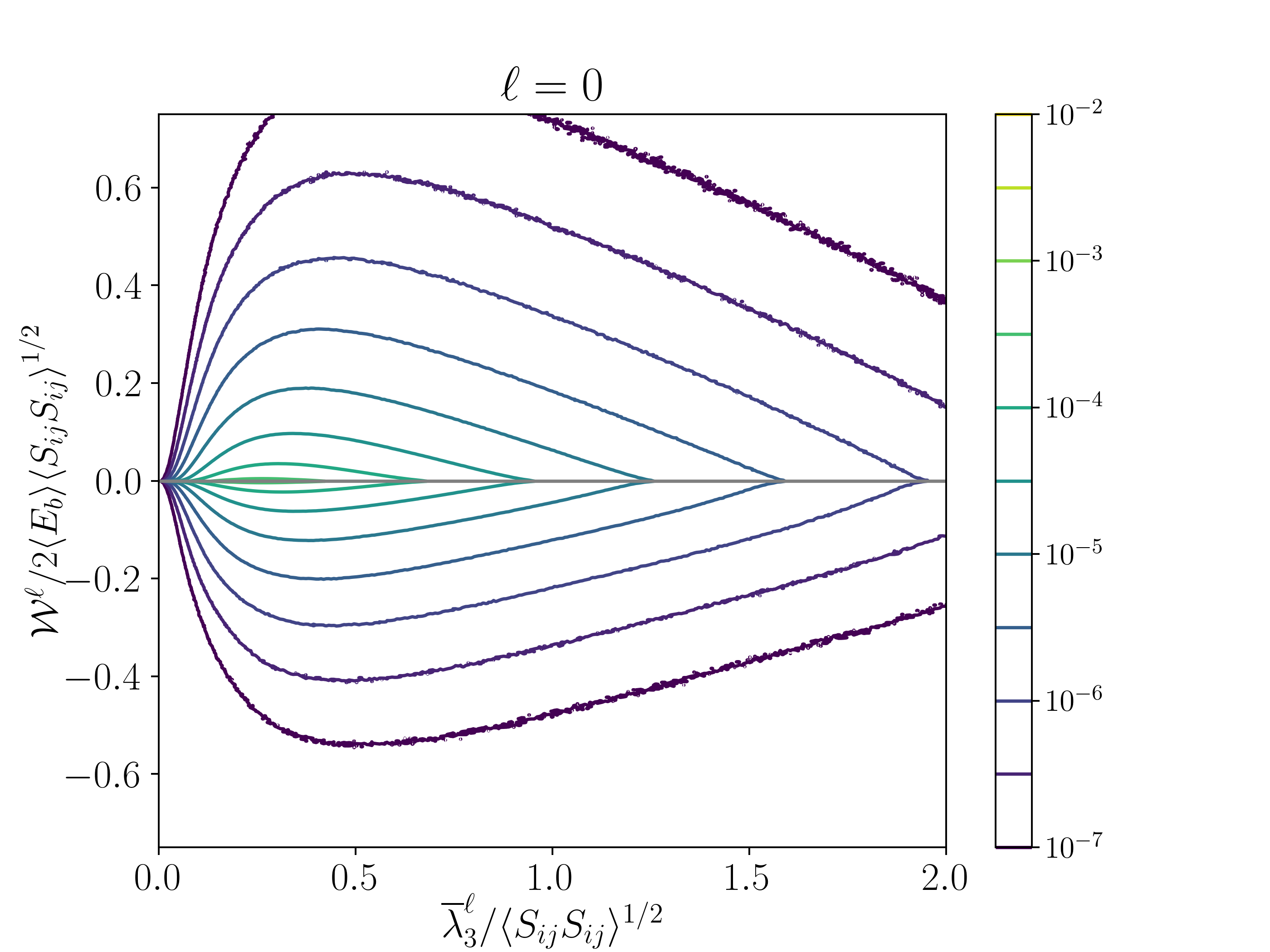} 
    \hspace{-1.5cm}
    \includegraphics[width=0.5\columnwidth]{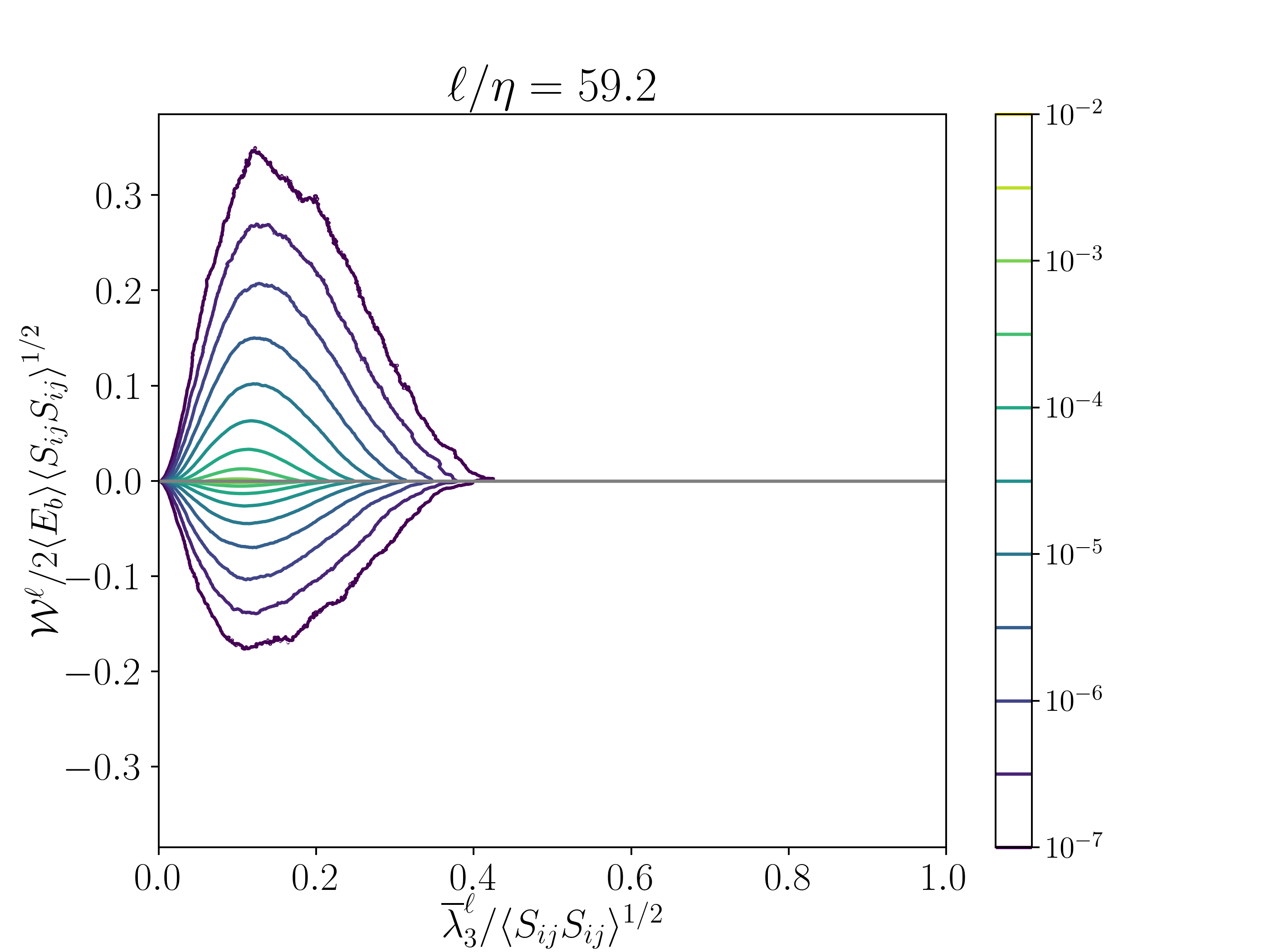} 
    
	 \caption{
Isolines of the joint pdf between $\mathcal{W}^\ell$ and the sign-definite strain-rate tensor eigenvalues. Top panels: $P({\mathcal{W}}^\ell,\overline{\lambda}^\ell_1)$. Bottom panels: $P({\mathcal{W}}^\ell,\overline{\lambda}^\ell_3)$.
In both cases, the left panels correspond to the full field, while the right panels correspond to the Inertial range. All the panels share the same aspect ratio.}
\label{fig:joint_W_first_third_eig}
\end{figure}

In this Appendix, we extend the discussion of sec.~\ref{sec:results_sheets} regarding the joint statistics between the energy conversion term and its degrees of freedom from eq.~\eqref{eq:conv_spec_filter}. 
%The analysis presented here, together with that partially discussed in sec.~\ref{sec:results_sheets}, was repeated on a lower-resolution dataset to verify the convergence of the observed trends. 
In particular, we start from $P({\mathcal{W}}^\ell,\overline{\lambda}^\ell_1)$ and $P({\mathcal{W}}^\ell,\overline{\lambda}^\ell_3)$, corresponding to the sign-definite strain eigenvalues.
%, as the joint statistics including the intermediate eigenvalue $\overline{\lambda}^\ell_2$ is already displayed in sec.~\ref{sec:results_sheets}. 
The inspection of the panels in fig.~\ref{fig:joint_W_first_third_eig} leads to the same conclusion: large energy-conversion events are not associated with regions of intense strain motion. This conclusion holds for both joint distributions and at both filtering scales shown. A common feature of all these cases is that the isolines are biased towards positive $y-$axis values as a consequence of the energy balance, i.e., $\lan \mathcal{W}^\ell \ran > 0 $.

\begin{figure}
	\centering
    \noindent\makebox[\textwidth]{
    \includegraphics[width=0.5\columnwidth]{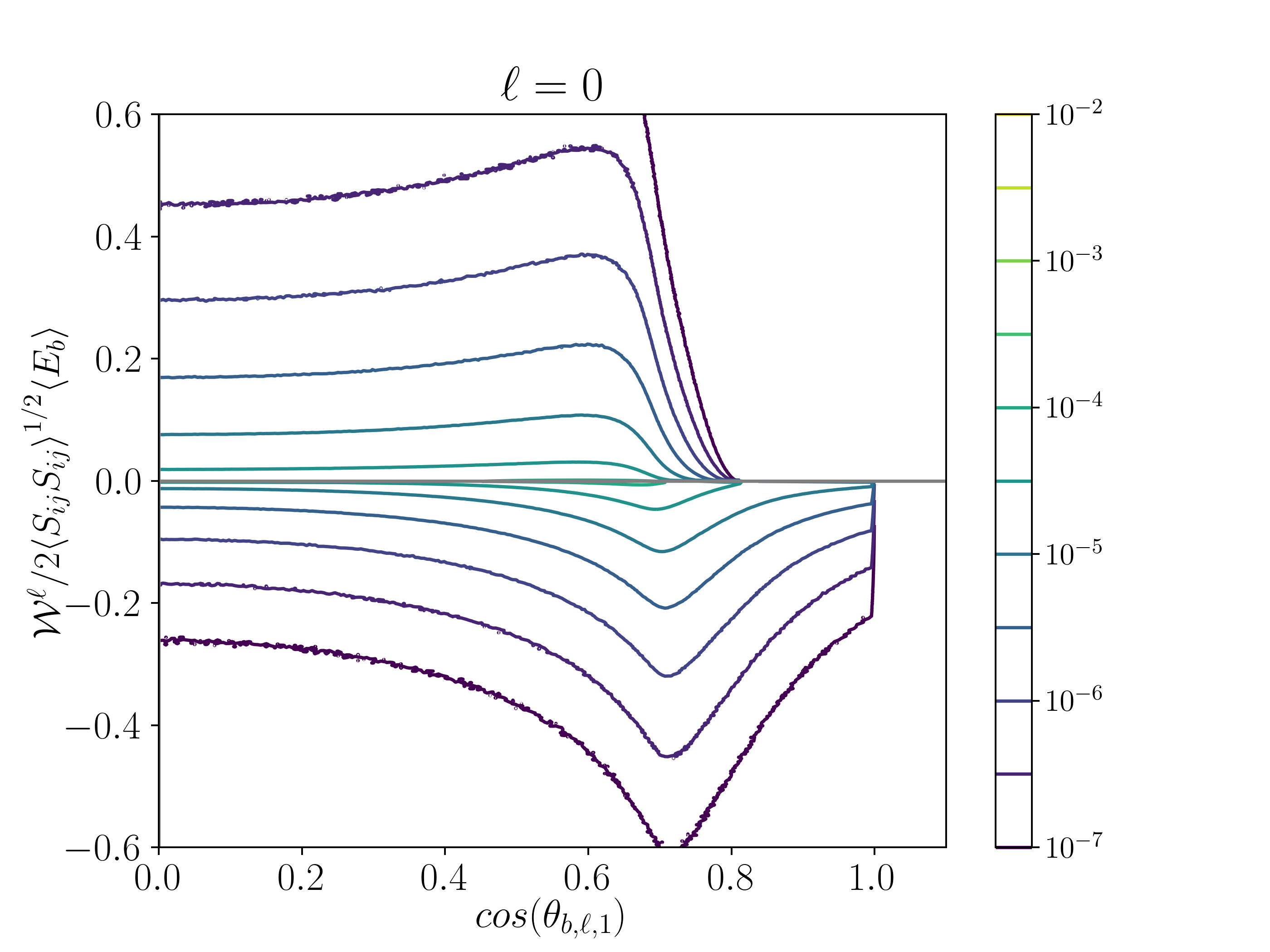} 
    \hspace{-1.5cm}
    \includegraphics[width=0.5\columnwidth]{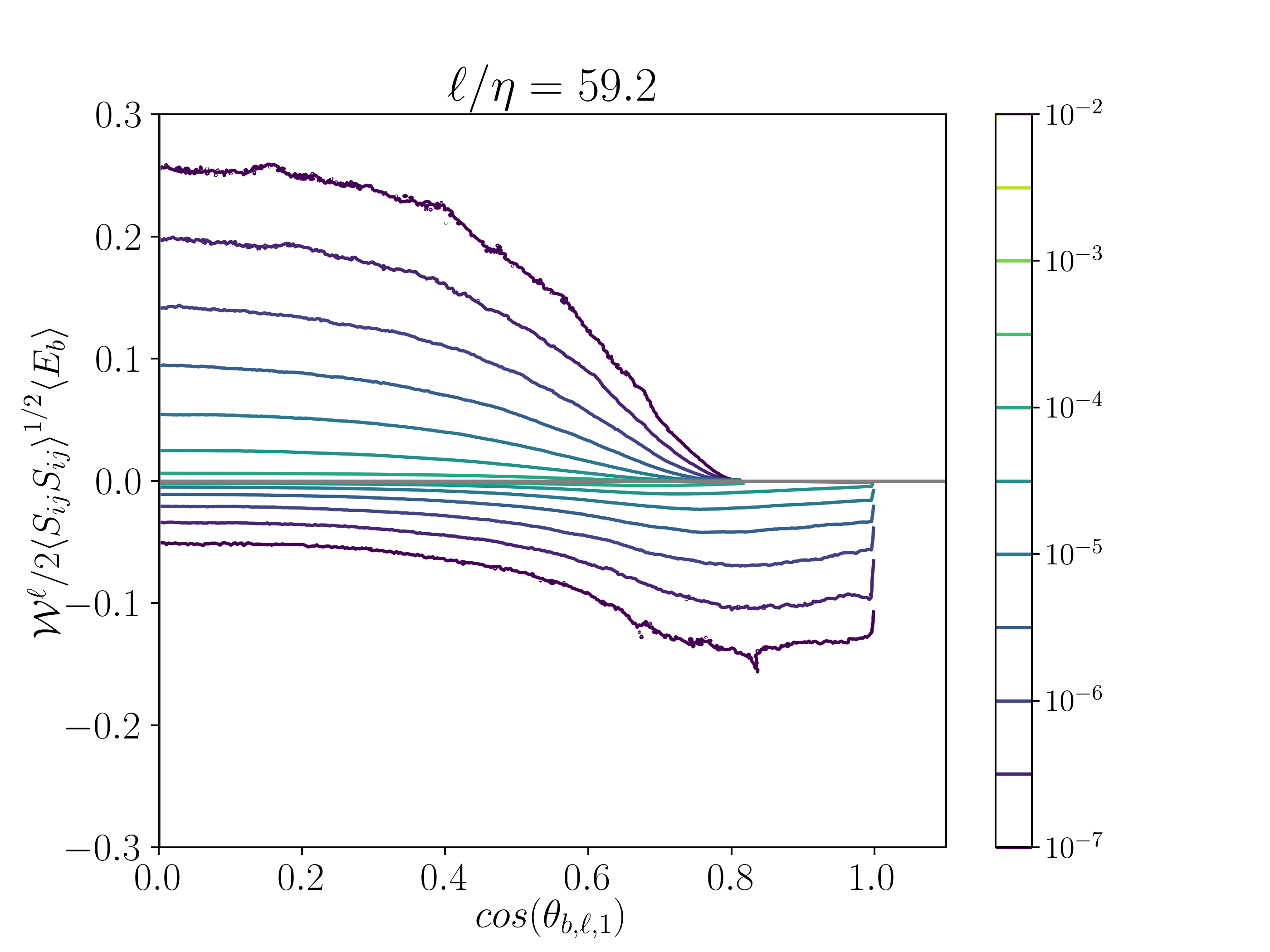}
    }
    
    \includegraphics[width=0.5\columnwidth]{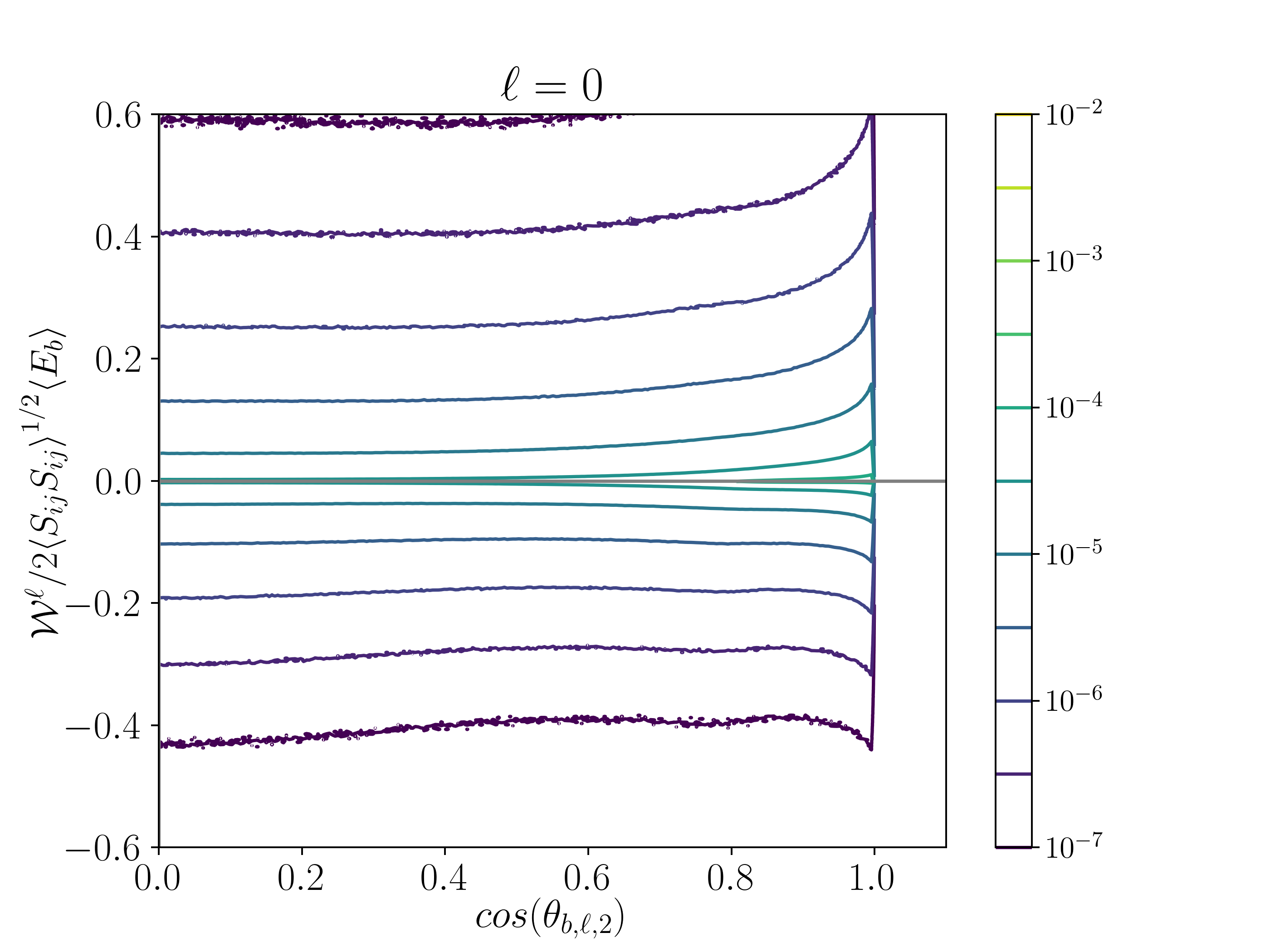} 
    \hspace{-1.5cm}
    \includegraphics[width=0.5\columnwidth]{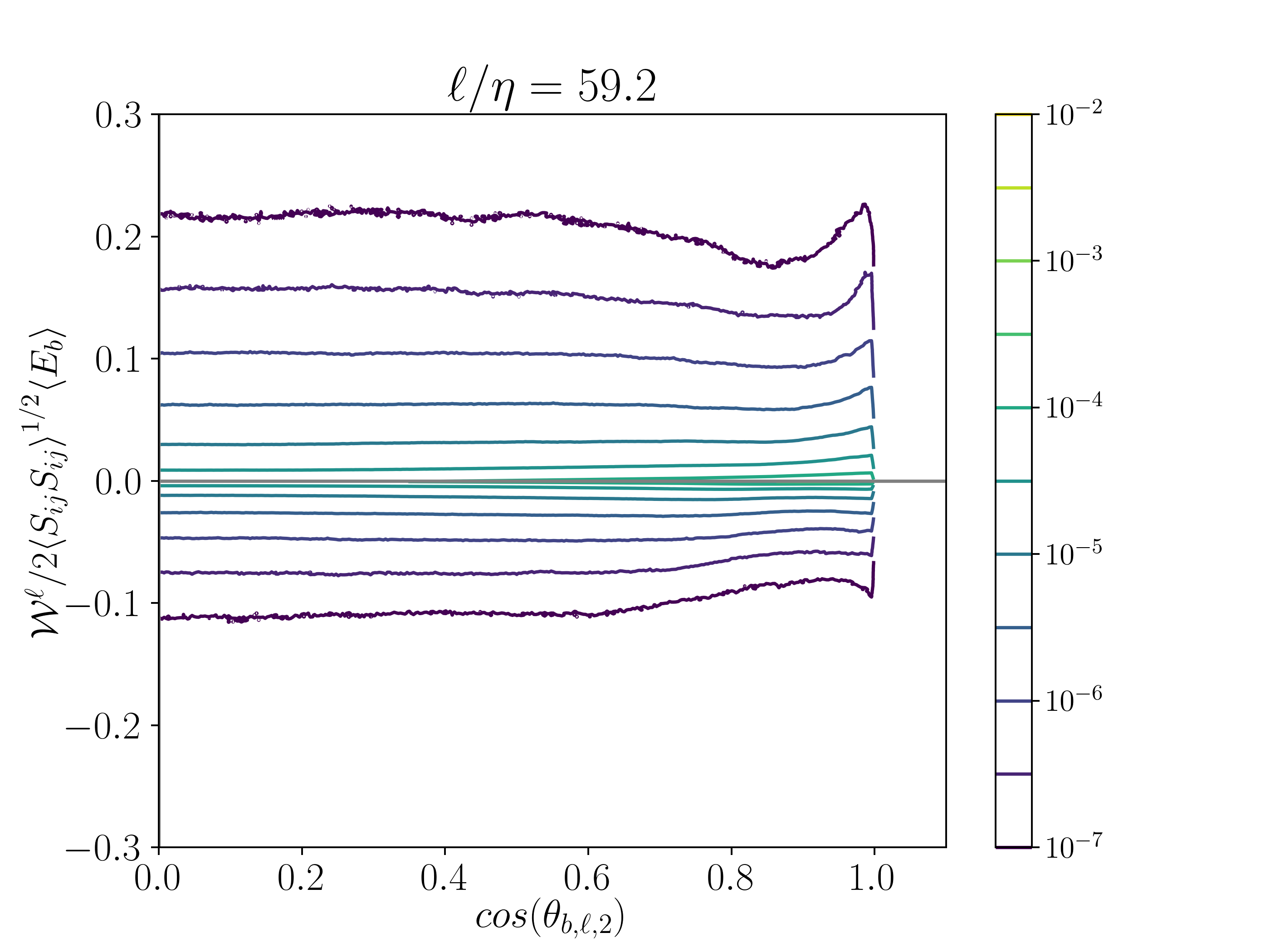} 
    
    \includegraphics[width=0.5\columnwidth]{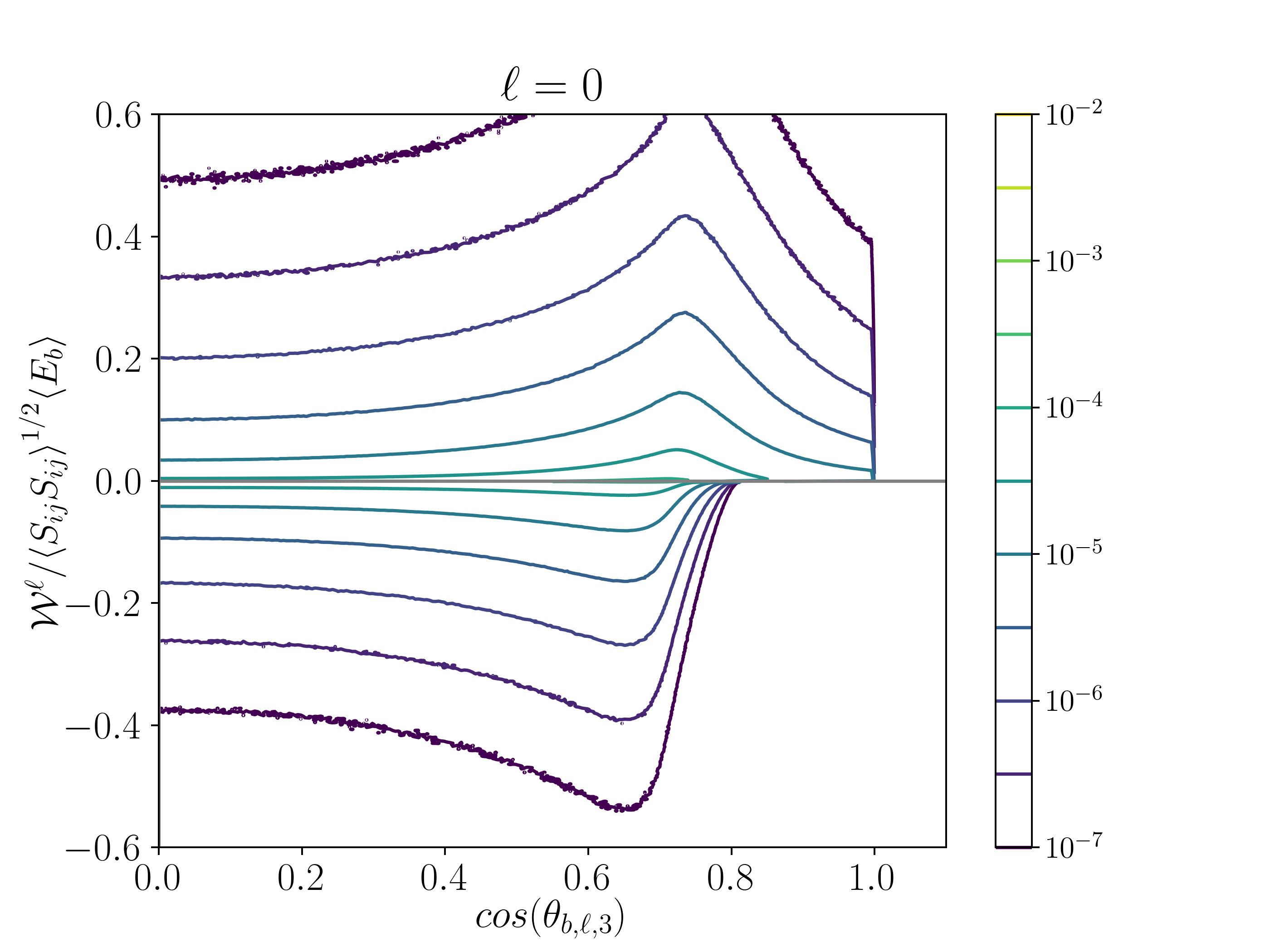} 
    \hspace{-1.5cm}
    \includegraphics[width=0.5\columnwidth]{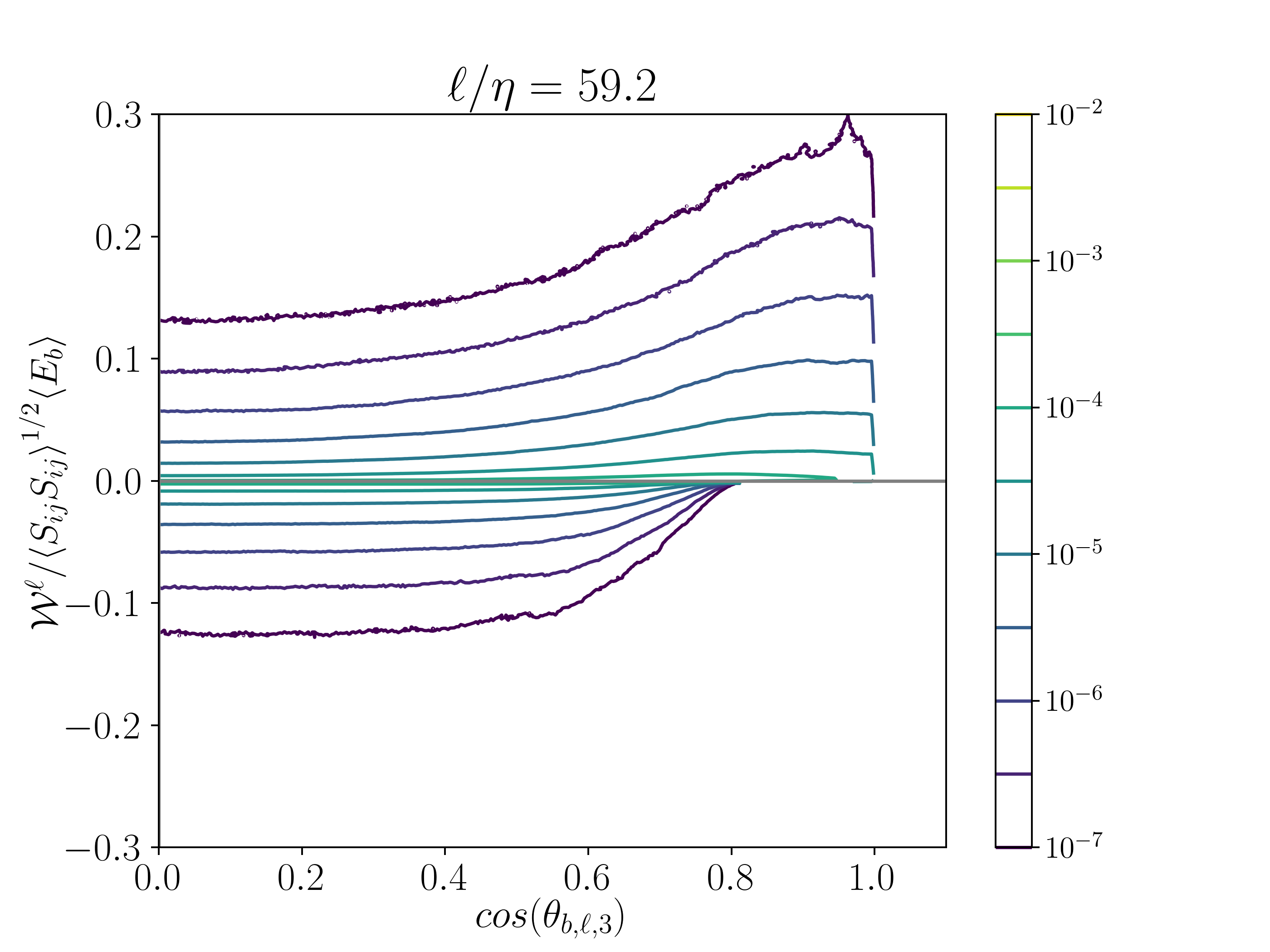} 
	 \caption{
Isolines of the joint pdf between $\mathcal{W}^\ell$ and the cosine of the angle between the strain eigenvectors and the magnetic field. The top, middle and bottom panels display respectively $P({\mathcal{W}}^\ell, \cos{\theta_{\bm{b},\ell,1}} )$, $P({\mathcal{W}}^\ell, \cos{\theta_{\bm{b},\ell,2}} )$ and $P({\mathcal{W}}^\ell, \cos{\theta_{\bm{b},\ell,3}} )$. 
In each row, the left panels correspond to the unfiltered fields, while the right panels correspond to the Inertial range. All the panels share the same aspect ratio. }
\label{fig:joint_W_cos}
\end{figure}

In fig.~\ref{fig:joint_W_cos}, we provide a quantitative description of the strain--magnetic-field alignment statistics discussed in sec.~\ref{sec:results_sheets}, focusing on the angles entering eq.~\eqref{eq:conv_spec_filter}. Starting with the eigenvector associated with $\overline{\lambda}_1^\ell$, at $\ell=0$ the positive conversion isolines reach their largest values of $\mathcal W^\ell$ at $\theta_{\bm{b},\ell,1}\simeq42^\circ$. For $\theta_{\bm{b},\ell,1}>42^\circ$, the isolines are nearly flat, whereas positive energy conversion is practically absent for $\theta_{\bm{b},\ell,1}<42^\circ$. This behaviour indicates only a modest alignment between the magnetic field and the eigenvector associated with the most compressive strain eigenvalue. As the filtering scale increases, this preferential orientation shifts progressively towards orthogonality.

By contrast, the behaviour associated with the eigenvector $\overline{\lambda}_3^\ell$ is approximately the opposite. Indeed, the corresponding panels are nearly related by a $y\to -y$ symmetry. For the unfiltered fields, the positive-conversion contours reach their largest values of $\mathcal{W}^\ell$ at $\theta_{\bm{b},\ell,3}\gtrsim42^\circ$, while their support extends down to $\theta_{\bm{b},\ell,3}=0^\circ$. As the filtering scale increases, this alignment becomes progressively stronger, with the largest positive-conversion values eventually occurring at $\theta_{\bm{b},\ell,3}=0^\circ$.

The strong scale dependence observed in the previous two cases is slightly less pronounced for the eigenvector associated with $\overline{\lambda}_2^\ell$. Large positive conversion is preferentially associated with strong alignment of the magnetic field with this direction, $\theta_{\bm{b},\ell,2}\simeq0$. This association persists as the filtering scale increases, although their angular dependence becomes progressively weaker.

\begin{figure}
	\centering
    \noindent\makebox[\textwidth]{
    \includegraphics[width=0.49\columnwidth]{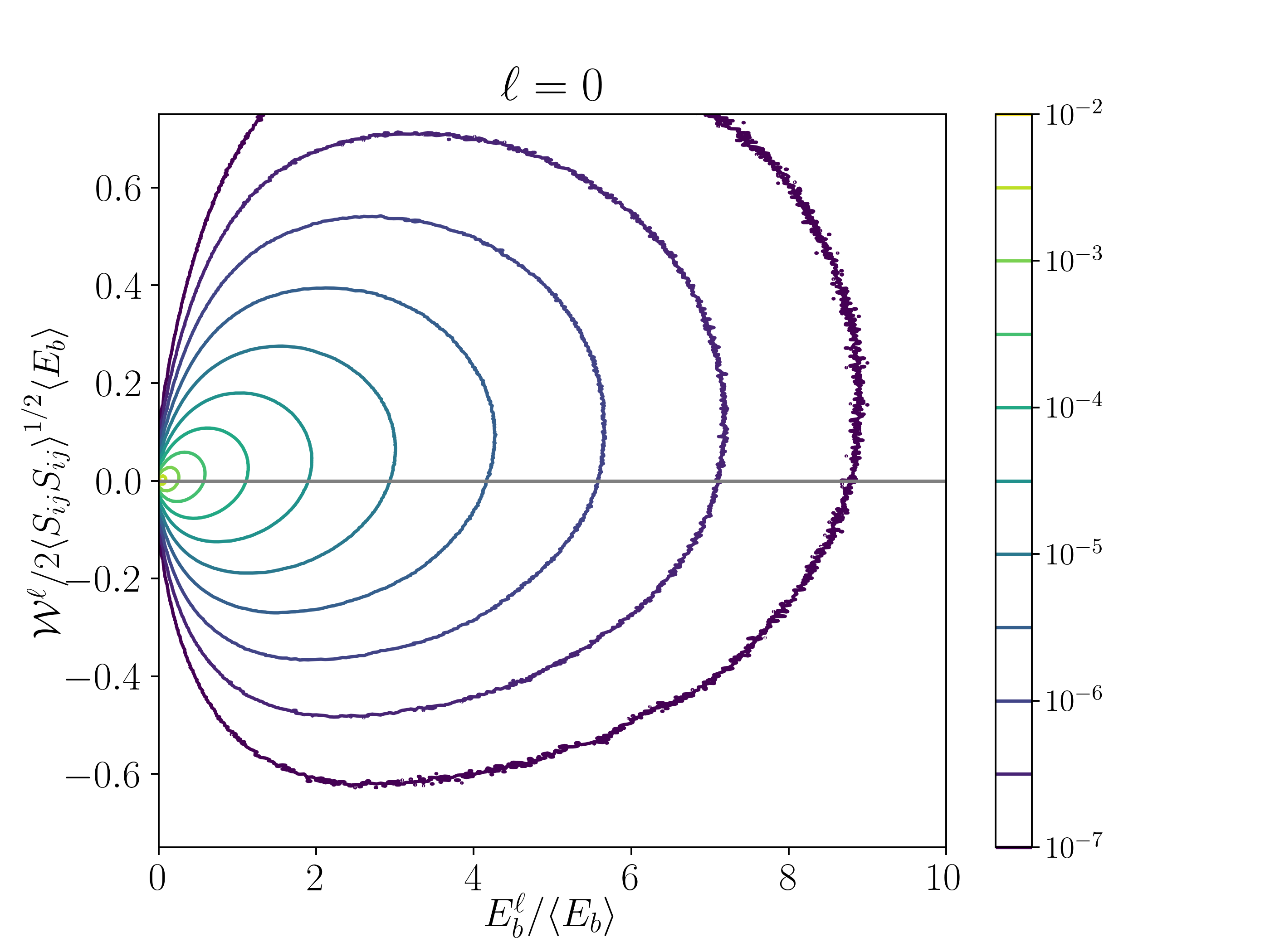} 
    \hspace{-1.5cm}
    \includegraphics[width=0.49\columnwidth]{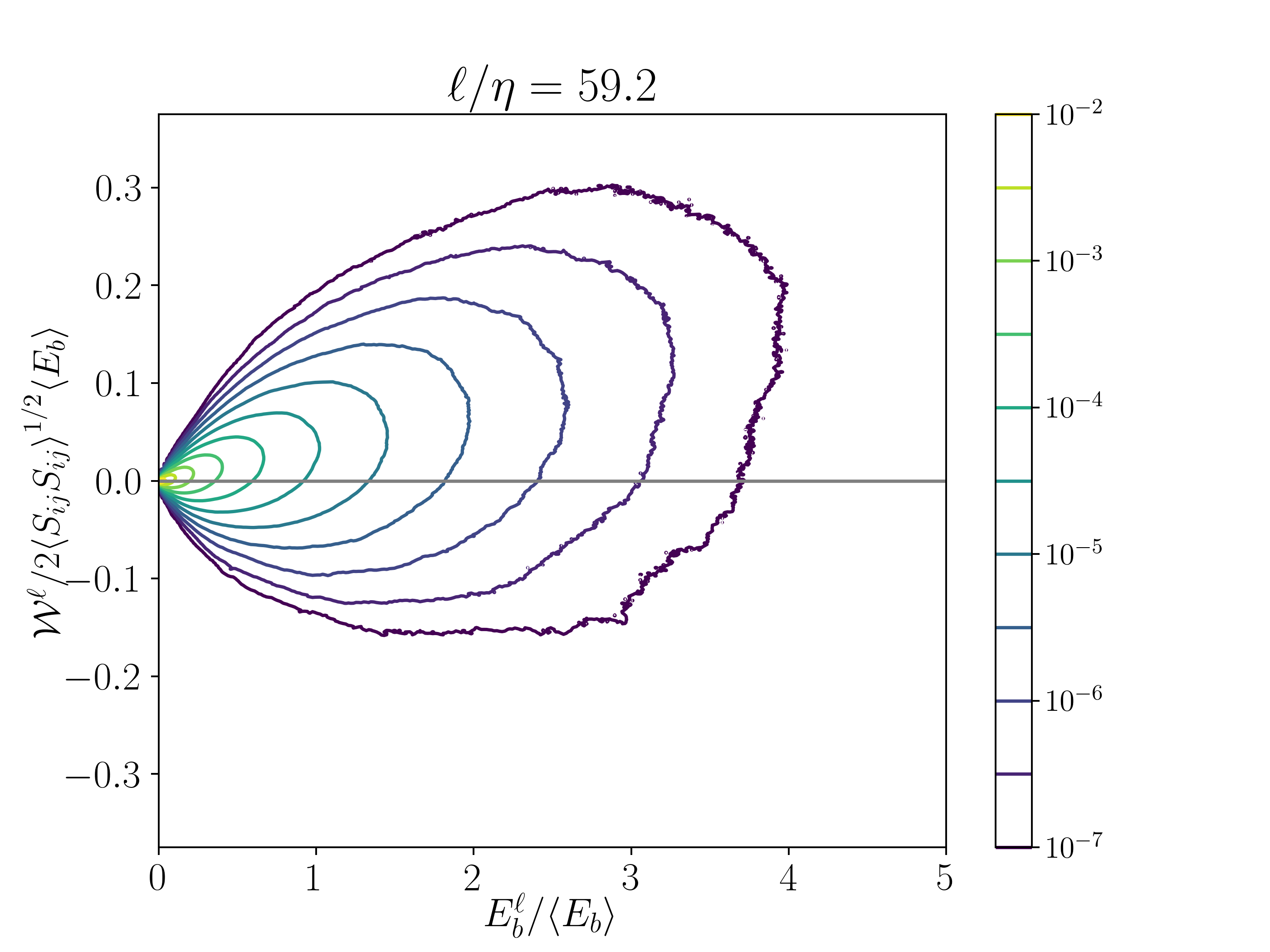} 
    }
	 \caption{
Isolines of the joint pdf between $\mathcal{W}^\ell$ and the resolved-scale magnetic energy $E_b^\ell = |\overline{\bm b}^\ell|^2/2$, corresponding to the full field (left panel) and to the Inertial range (right panel). The aspect ratio is the same in both panels.}
\label{fig:joint_W_mag}
\end{figure}

Concerning the joint statistics with the magnetic energy, namely $P(\mathcal{W}^\ell,|\overline{\bm{b}}^\ell|^2)$, displayed in fig.~\ref{fig:joint_W_mag}, we observe no clear association between large energy conversion and regions of strong magnetic field. Indeed, the isolines are approximately elliptical, indicating only a weak correlation between the two variables. The only noticeable asymmetry is that between positive and negative conversion, which is simply a consequence of the energy balance, as already discussed for the other observables.

Finally, we consider the joint statistics $P(\mathcal{W}^\ell,\overline{\lambda}^\ell_i\cos^2\theta_{\bm{b},\ell,i})$, where the second argument corresponds to the right-hand side of eq.~\eqref{eq:conv_spec_filter} normalised by $|\overline{\bm{b}}^\ell|^2$. This isolates the combined role of strain intensity and strain--magnetic-field alignment from that of the magnetic-field amplitude, whose association with the conversion is weak, as shown in fig.~\ref{fig:joint_W_mag}. This normalisation is also relevant experimentally, since, as discussed in sec.~\ref{sec:experiments}, spacecraft studies commonly examine energy conversion after normalising by the magnetic-field strength squared. We find that large positive energy conversion is associated with moderate values of $\overline{\lambda}^\ell_i\cos^2\theta_{\bm{b},\ell,i}$, whose characteristic magnitude displays a surprisingly weak dependence across the scales (not shown).

\clearpage

\bibliographystyle{plain}
\bibliography{thesis}

%% \end{document} %.. to drop the revtex advice/info sections
%===========================================================

\end{document}